\documentclass[journal=nalefd,manuscript=article]{achemso}

\usepackage[version=3]{mhchem} 

\author{Ilya Golokolenov}
\affiliation[Institut Neel]
{Univ. Grenoble Alpes, Institut N\'eel, 25 rue des Martyrs,
Grenoble, 38042, France}
\author{Sumit Kumar}
\affiliation[Institut Neel]
{Univ. Grenoble Alpes, Institut N\'eel, 25 rue des Martyrs,
Grenoble, 38042, France}
\alsoaffiliation[RHUL]
{Royal Holloway University of London, Egham, Surrey, TW20 0EX, England}
\author{Baptiste Alperin}
\affiliation[Institut Neel]
{Univ. Grenoble Alpes, Institut N\'eel, 25 rue des Martyrs,
Grenoble, 38042, France}
\author{Bruno Fernandez}
\affiliation[Institut Neel]
{Univ. Grenoble Alpes, Institut N\'eel, 25 rue des Martyrs,
Grenoble, 38042, France}
\author{Andrew Fefferman}
\affiliation[Institut Neel]
{Univ. Grenoble Alpes, Institut N\'eel, 25 rue des Martyrs,
Grenoble, 38042, France}
\author{Eddy Collin}
\affiliation[Institut Neel]
{Univ. Grenoble Alpes, Institut N\'eel, 25 rue des Martyrs,
Grenoble, 38042, France}
\email{eddy.collin@neel.cnrs.fr}
\title[An \textsf{achemso} demo]
  {Nano-beam clamping revisited}

\abbreviations{NEMS}
\keywords{Nano-mechanics (NEMS), mode frequency, mode dissipation, soft clamping}

\begin{document}


\begin{abstract}
 Within recent years, the field of nano-mechanics has diversified in a 
variety of applications, ranging from quantum information processing 
to biological molecules recognition. 
Among the diversity of devices produced these days, the simplest (but versatile) element 
remains 
the \textit{doubly-clamped beam}: it can store very large tensile stresses (producing high resonance frequencies $f_0$ and quality factors $Q$), is interfaceable with electric setups (by means of conductive layers), and can be produced 
easily in clean rooms (with scalable designs including multiplexing).
Besides, 
its mechanical properties are the simplest to describe.
Resonance frequencies and $Q$s are being modeled, with as specific achievement the ultra-high quality resonances based on ``soft clamping'' and ``phonon shields''. 
Here, we demonstrate that the fabrication undercut of the clamping regions of basic nano-beams produces a ``natural soft clamping'', given for free. We present the analytic theory that enables to fit experimental data, which can be 
used for $\{ Q , f_0 \}$ design:
beyond Finite Element Modeling validation, the presented expressions provide a profound understanding of the phenomenon, with both a $Q$ enhancement and a downwards frequency shift.
\end{abstract}


\section{Introduction}

Doubly-clamped nano-beams are utterly basic, but nonetheless remarkably versatile.
They are routinely used in a broad range of applications, from mass sensing\cite{roukes_mass} to quantum electronics\cite{fink_circu}.
About two decades ago, it had been found that stoichiometric Silicon-Nitride (Si$_3$N$_4$) thin films grown on Silicon can store very large tensile stresses, and that subsequently mechanical nano-structures patterned on this material display very high flexural resonance frequencies $f_0$, and interestingly \textit{very high $Q$ factors}\cite{jeevak_APL}. 
This phenomenon has been named ``dissipation dilution'': it is due to the large stored elastic energy, as compared to the losses which arise from bending\cite{Jeevak_NanoLett,PRB_kippen}.
Indeed, bending losses have been found experimentally to be essentially stress-independent\cite{VillanuevaSchmidPRL,FtouniPRB,martial_PHD}.
The precise amount of stress stored in the structures can be tuned by stoichiometry, but also by chip-bending\cite{Jeevak_NanoLett}, by the design of the beam shape\cite{nano_poot} and by the clamping pillars\cite{stress_weig}.

A precise (yet phenomenological) modeling of the flexure of beams had been proposed considering ideal clamping\cite{prl_unterrheitmeier}, and then successfully adapted to membranes\cite{prl_regal}.
It is based on a numerical solution of the \textit{Euler-Bernoulli} equation which describes the low-frequency dynamics of thin-and-long beams\cite{cleland_book}, matching the observed $Q$ factor linear increase with beam length $L$, and decrease with mode number $n$.
The key argument is to assume that internal microscopic friction mechanisms originate from the bending of the material, whatever they might be.
In this sense, the model applies as well to pure nitride structures\cite{prl_unterrheitmeier} as to bi-material devices where a metallic layer dominates the damping\cite{jeevak_metal, prl_regal, eddy_JAP, martial_PHD}. It also works for both room-temperature\cite{prl_unterrheitmeier, prl_regal} and low temperature\cite{martial_PHD} experiments. For the former, it has been argued that losses are dominated by surface effects\cite{VillanuevaSchmidPRL}; for the latter, a specific mechanism based on Two-Level-Systems (TLSs) present in the materials is discussed in the literature\cite{davisTLS,kunal_PRL,OliveArXiv}.

Beyond internal damping, the anchoring points appear to play a very important role in beam dynamics. 
The vibration of the mechanical modes irradiates acoustic waves in the supports, which limits the $Q$. This radiation damping has been modeled in particular for thin supports\cite{judge1,judge2,crossPRB,ignacio}, which is typically the geometry obtained when the fabrication process under-etches the clamps (so-called \textit{undercut}).
Such structures are very common in the literature,  and are the focus of the present manuscript.
The actual limiting $Q$-value depends on the precise geometry of the beam (particularly its width $w$ and length $L$), and of the anchor. Experimentally, for 
the first flexure $n=1$ of 
millimeter length low-stress structures resonating at sub-MHz frequencies, 
radiation loss seems irrelevant for widths $w<3~\mu$m\cite{schmidPRB2011};
for high-stress devices (tens of MHz frequencies) with width $w \sim 200~$nm, acoustic radiation dominates for $L<10~\mu$m\cite{kippNanoLett}.
This leaves a 
large playground to experimentalists where bending is the main source of losses\cite{JeevakMaxQ}. On the other hand, 
when radiation losses dominate it is possible to suppress them with a clamp structuring that forbids phonon transport at the specific frequency of the mode $n$ that one wants to protect: a so-called ``phonon shield''\cite{kippNanoLett, schliesser_Q, RegalAPL}.

But the anchoring does more than enabling irradiation into the bulk: it defines the \textit{precise bending shape} at the clamping point.
It had been realized about a decade ago that most of the friction occurs near the beam's ends, where the bending is the most dramatic\cite{schmidPRB2011,SuhelPRB}. 
As such, beyond creating phonon gaps in the substrate's density of states, structuring the clamps has another (more trivial) effect: it can reduce this bending and mitigate the losses, which is named ``soft clamping''\cite{schliesser_Q,kipp_science,schmidJAProundclamp}. 
In this Letter, we demonstrate experimentally that a clamp undercut acts essentially as a ``soft clamping'' given for free: the quality factor $Q$ of the flexural modes \textit{grows} with the beam width $w$, which is the reverse behavior when compared to acoustic radiation. As well, the clamp modifies substantially the resonance frequencies\cite{VdZ_freq} (we show that it \textit{decreases} with $w$), which can be used for design purposes, for instance for nano-beam resonance multiplexing\cite{Ilya_JLTP}.

Early beam clamp modeling relied on phenomenological ansatzs\cite{schmidPRB2011,SuhelPRB}. Instead, here we present an exact analytic theory which follows the same lines as the modeling performed on membranes\cite{prl_regal}.
It is based on a high-stress Taylor expansion of the modal parameters (for any mode $n$), introducing as fit parameters a mass-loaded spring and torque at each of the beam's ends.
We demonstrate that \textit{both} the resonance frequency $f_0$ and $Q$ can be fit with a \textit{clamp spring coefficient} $\alpha_{l,r}$ ($l,r$ for left and right), the torque being (at lowest order) negligible. 
$\alpha_{r,l}$ is found to be proportional to frequency and inversely proportional to width $w$, the prefactor being a characteristic of the anchor's shape (and material) solely. 
We believe that our results constitute a very useful tool when designing \textit{basic} doubly-clamped beams presenting a characteristic fabrication undercut.

\section{Results}

\vspace{0.2cm}
\begin{figure}
\includegraphics[width=\linewidth]{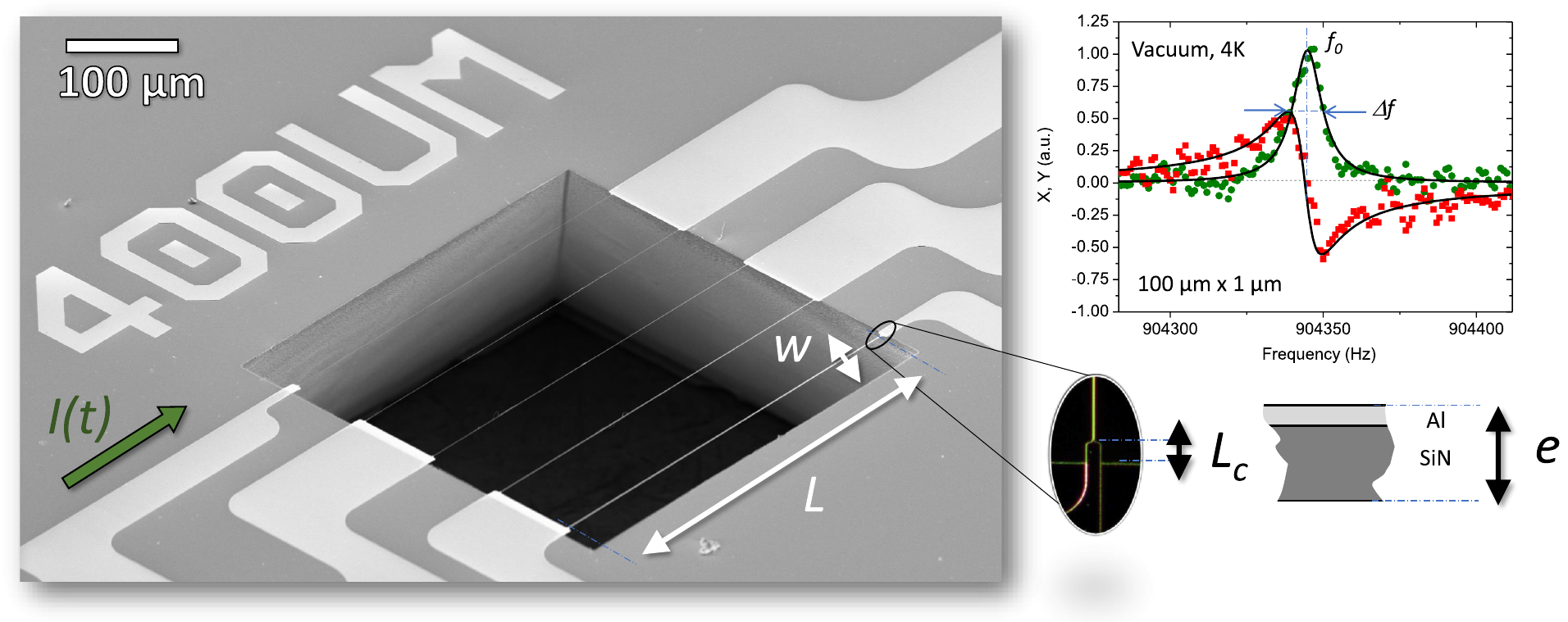}
\caption{\label{figure1} Left: SEM image of one sample ($400~\mu$m long beams), with device widths $w$ of 50$~$nm, 100$~$nm, 200$~$nm, 500$~$nm and 1000$~$nm (from top to bottom). Bottom-right inset: dark field image of clamping region (suspended part of length $L_c$), and schematic of the layered structure (not to scale). Top-right: typical phase-resolved linear response line measured for one of our devices (first flexure $n=1$ of a 100$~\mu$m long, 1$~\mu$m large one). Drive current $I_0$ of 0.5$~$nA, $B_0$ field 0.2$~$T. Lines are Lorentzian fits with normalized heights (see text for details). }
\end{figure}

\subsection{Experimental details}
Nano-beams of various widths and lengths have been realized from low-stress Silicon-Nitride. All have a thickness of about 100$~$nm, and come from the same wafer.
A conductive layer (30$~$nm Aluminum) has been deposited on top in order to create electrical contacts (total thickness $e=130~$nm).
A typical sample SEM image is shown in Fig. \ref{figure1} (left), with a zoom-in on one of the clamping regions (inset, with schematic of the 
bilayer 
structure).
The design is such that the beams are all \textit{fully suspended} within a hollow window; they are connected to the bulk only through a well-defined over-hanging clamp of length $L_c$ (of order $12~\mu$m for all of them). 
This part has the same thickness (and same constitution) as the rest of the beam, which guarantees that all the layer's stress is transmitted within the structure. 
The absence of pedestals holding the beams 
enables to avoid stress-relaxation effects occurring when beams are released, and the pedestals  bending\cite{stress_weig}. 
The over-hanging rectangular zone is actually defined through the wet KOH etch: due to the 54.74$^\circ$ angle that the etching creates in the silicon substrate (referenced to the wafer surface), the opening is smaller on the front side  than on the back. This generates very straight and clean suspended clamp regions, ideal for a model experiment. It has to be contrasted with usual undercuts linked to the releasing etching time, which are less well-defined \cite{VdZ_freq}. 
A description of the fabrication process can be found in Ref.\cite{Ilya_JLTP}. Details about sample characteristics can be found in Supplementary Information.

The measurements are performed using the magnetomotive technique\cite{roukesclelandsensors}. A current $I(t)=I_0 \cos(\omega t+\varphi)$ is fed into the beam which stands in a static magnetic field $B_0$, thus generating a force $\propto I_0 B_0$ at frequency $\omega$. 
Used fields typically range from 0.2$~$T to 1$~$T, and care is taken to take all data in the linear regime, characterizing carefully the extra damping coming from the electric circuit\cite{roukesclelandsensors} (which needs to be subtracted). 
Each sample consists in a set of beams of 
equal lengths, but with widths ranging from 50$~$nm to 1000$~$nm. For a given sample, all beams can be connected in series, which enables a straightforward multiplexing (see Fig. \ref{figure1}). 
Besides, each beam can also be connected independently, in order to separate its resonances from other devices without ambiguity. 
As beams move out-of-plane, a voltage $V(t)$ is induced at their extremities, proportional to field $B_0$ and velocity. Only symmetric modes are detected, the signal being proportional to the maximal amplitude of motion; anti-symmetric ones cancel out.
Measurements are conducted in cryogenic vacuum at 4.2$~$K, using a lock-in amplifier (with in-phase $X$ and quadrature $Y$ components). An example of resonance peaks obtained via a frequency-sweep is shown in Fig. \ref{figure1} (right). More details on the setup can be found in Ref\cite{Ilya_JLTP}. \\

\subsection{Data Analysis and Theory}
From the measured peaks one extracts 
resonance 
frequency $f_0$ (centre position) and damping $\Delta f$ (full width at half height on $X$ component), giving us the quality factor $Q=f_0/\Delta f$. This is performed on three different 100$~\mu$m long samples, two sets of 300$~\mu$m and 400$~\mu$m ones, and one set of 200$~\mu$m beams. Only on one set (100$~\mu$m long, 1$~\mu$m 
wide) did we measure 
the mode dependence (with $n=1,3,5$). Only very few points have been dropped from the statistics, due we believe to fabrication irreproducibility: the 50$~$nm beams are not very homogeneous in width, and even break with thermal cycling. On some beams, defects can be seen (which look like some sort of filaments, see Fig. \ref{figure1}), which might not be negligible for the 
narrowest $w=100~$nm devices.  
Besides, we see a similar scatter in measured damping (up to a factor 2) as in Ref\cite{prl_unterrheitmeier}. from one sample to the other. The final error bars in our graphs therefore reflect the statistical scatter due to the fabrication process. For details, see Ref\cite{Ilya_JLTP}. and Supplementary Information. 

\vspace{0.2cm}
\begin{figure}
\includegraphics[width=\linewidth]{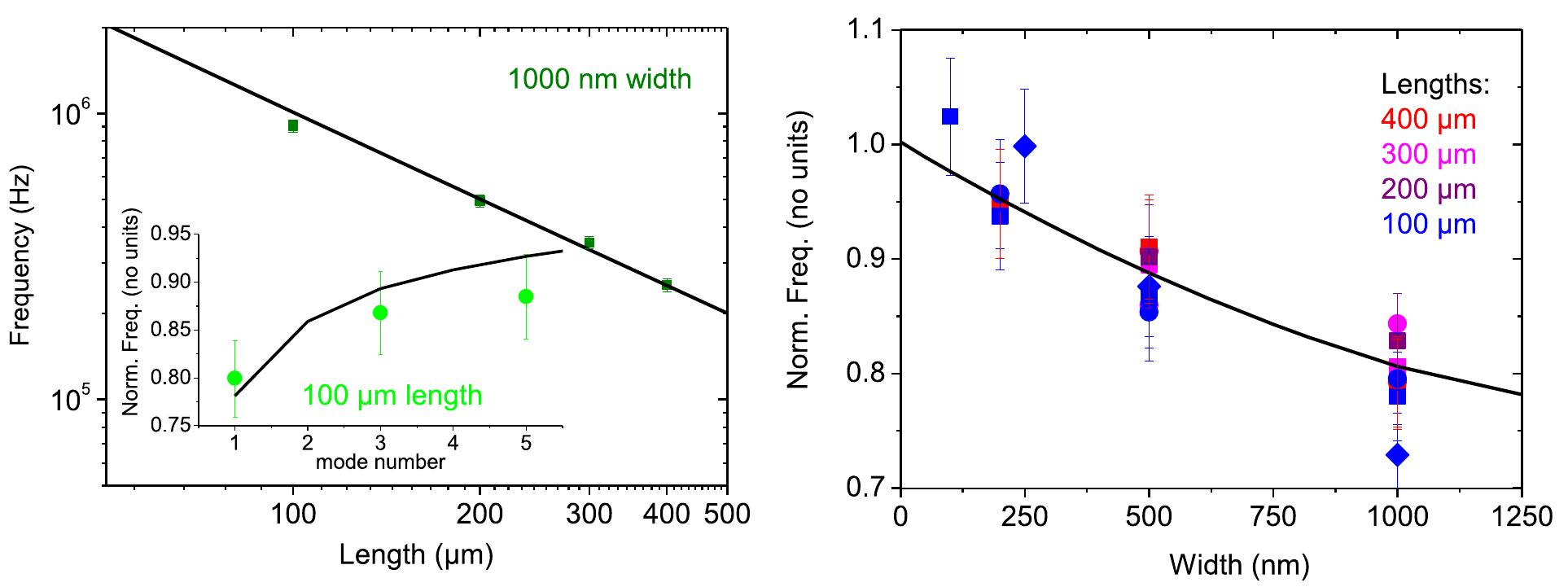}
\caption{\label{figure2} Left, main: 
dependence of frequency on 
length $L$ for the first flexure $n=1$ (all devices having $w=1000~$nm). Left, inset: frequency of 100$~\mu$m long, $1000~$nm wide device normalized to the extrapolated value $f_{n,0}$ 
at $w \rightarrow 0$ 
as a function of mode number $n$. Right: same normalized frequency for mode $n=1$ of all devices, as a function of width $w$. The lines correspond to Eqs. (\ref{fitf},\ref{freq}) with $L=100~\mu$m, using a simple ansatz for the clamp parameter $\alpha$, see text.  }
\end{figure}

The starting point of the modeling is the Euler-Bernoulli equation that describes the flexure of long-and-thin beams\cite{cleland_book}. As in Ref\cite{prl_regal}. we define as small parameter:
\begin{equation}
a = \sqrt{\frac{E \, I_z}{S L^2}} , 
\end{equation}
with $E$ the Young's modulus and $S=\sigma \, w e$ the tensile force acting on the beam ($\sigma$ is the in-built axial stress).
The second moment of area $I_z=\frac{1}{12} w e^3$ is a geometrical parameter. The composite nature of the beam can be accounted for by choosing an effective Young's modulus $E$ that depends on the $E_{SiN}$ and $E_{Al}$ of the two materials and on their thickness\cite{timoshenko} (and taking for $\rho$ the mean density). 
In practice, this parameter is close to the $E_{SiN}$ elastic constant, about 200$~$GPa ($\pm 50~\%$) from the literature\cite{prl_unterrheitmeier,stress_weig}. Numerical values for our experiment can be found in Supplementary Information.
The specificity of the approach lies in the boundary conditions: on each side ($l,r$ for left and right) we impose an elastic force and an inertial force. They are combined into an \textit{effective spring constant} $k_{l,r}=\frac{S}{L} \alpha_{l,r}$. Similarly, a torque (that combines elastic and inertial contributions) acts also on each end of the beam, resulting in an \textit{effective torsion spring} $\Gamma_{l,r}= S L \, \gamma_{l,r}$.
The (dimensionless) 
constants $\alpha_l, \gamma_l$ and $\alpha_r, \gamma_r$ fully characterize each of the $l, r$ clamps, respectively.
Ideal clamping is recovered with $\alpha_{l,r}, \gamma_{l,r} \rightarrow \infty$, which leads to the conventional boundary conditions (no displacement, no bending angle at both ends).
The exact analytical solutions for $f_0$ and $Q$ are finally derived to second order in $a$, and $1/\alpha_{l,r}, 1/\gamma_{l,r}$ (the explicit damping model producing $\Delta f$ is discussed thereafter).
We fit them on data, and find out that $\gamma_{l,r}$ can be to first approximation neglected; besides, only symmetric clamping will be addressed with $\alpha_{l}=\alpha_{r}=\alpha$. Nonetheless, the full mathematical description is given in Supplementary Information.

 Let us first discuss the resonance frequency parameter $f_0$ (given here in Hz).
For any mode $n$, it writes:
\begin{equation}
f_0(n,a,\alpha) = f_{n,0} \, P_f(n,a,\alpha) , \label{fitf}
\end{equation}
with $2 \pi \times f_{n,0}= \frac{n \pi}{L}\sqrt{\frac{\sigma}{\rho}}$ the usual $n^{ th }$ resonance frequency of a string (of density $\rho$), and $P_f$ a correction function. The latter is found to be:
\begin{eqnarray}
P_f(n,a,\alpha) & = & 1- \frac{2}{\alpha} + \frac{4}{\alpha^2} \nonumber \\
&& \!\!\!\!\!\!\!\!\!\!\!\!\!\!\!\!\!\!\!\!\!\!\!\!\!\!\!\!\!\!\!\!\!\!\!\!\!\!\!\!\!\!\!\!\!\!\! + \, a \left(2 - \frac{8}{\alpha} + \frac{2(12- n^2 \pi^2)}{\alpha^2} \right) \nonumber \\
&& \!\!\!\!\!\!\!\!\!\!\!\!\!\!\!\!\!\!\!\!\!\!\!\!\!\!\!\!\!\!\!\!\!\!\!\!\!\!\!\!\!\!\!\!\!\!\! + \frac{1}{2} \,a^2 \left(8+ n^2 \pi^2 - \frac{2(24+5 n^2 \pi^2)}{\alpha} + \frac{24(8+ n^2 \pi^2)}{\alpha^2}\right) \!  . \label{freq}
\end{eqnarray}
In Supplementary Information, we compare this result to the exact numerical Bokaian calculation\cite{bokaian} valid for ideal clamping ($\alpha \rightarrow \infty$). For $a<0.1$ the agreement is very good, but degrades with increasing $n$, demonstrating that higher orders need to be taken into account in the Taylor $a$-expansion.
Note the difference between Eq. (\ref{freq}) and expressions that can be found in the literature\cite{kippNanoLett,stress_weig}.

Experimental data and theory are compared in Fig. \ref{figure2}.
On the left panel, we present the expected $f_0$ scalings with $1/L$ (inverse length) and $n$ (mode number; in the normalized plot, $f_0(n)/f_{n,0}$ is at first order a constant).
However on the right panel, we demonstrate how the width $w$ of the beam influences the resonance frequencies:
$f_0(n=1)/f_{n=1,0}$ \textit{decreases} with increasing width, independently of length $L$. This is rather surprising (since $f_0$ is 
independent of $w$ in the high-stress limit of Euler-Bernoulli theory), 
and the effect is less pronounced for higher $n$ (inset left panel). Inspecting Eq. (\ref{freq}), we see that this behavior can be obtained by a simple ansatz on the clamp parameter $\alpha$:
\begin{equation}
\alpha \propto \frac{n \pi}{w} , \label{alphafit}   
\end{equation}
using the simplest guess. This actually means that the anchor becomes \textit{more stiff at higher frequencies, and for smaller beam widths}.
Eq. (\ref{alphafit}) is fit on data, see lines in Fig. \ref{figure2}, demonstrating very good agreement. 
The extracted parameter is characteristic of our clamp geometry, especially its length $L_c$. How the resonance frequency of cantilevers decreases with increasing $L_c$ had been investigated numerically in Refs\cite{VdZ_freq,Sadewasser}. Here, our approach is to fit this dependence by an effective spring constant $\alpha$; see Supplementary Information for quantitative parameters, and discussion in the Conclusion Section.

We now bring our attention to the quality factor $Q$, or equivalently the damping parameter $\Delta f=f_0/Q$. We remind that for our devices, radiation loss can be safely neglected\cite{kippNanoLett}.
Beyond the nanomechanics literature which clearly established that friction is directly related to bending\cite{prl_unterrheitmeier,prl_regal}, we shall discuss internal damping from a \textit{materials science} perspective. 
An ideal solid described in continuum mechanics obeys elasticity theory\cite{cleland_book}: one introduces strain and stress tensors which are related linearly by two elastic constants (for isotropic materials), e.g. Young's modulus $E$ and Poisson's ratio $\nu$. When friction mechanisms take place, deviations from this ``Hookean'' behavior appear: the solid is called \textit{anelastic}. 
This essentially means that one has to introduce a non-conservative force which acts upon each elementary volume $\delta \tau$ of the material.
The most natural modeling consists in introducing the rates of change of strain and stress tensors in the elastic equations, while keeping the hypothesis of linearity\cite{zenerbook}.

The simplest such linear superposition is 
the \textit{Zener model}\cite{zenerbook}, and it describes rather well low-frequency (kHz) elastic properties of conventional materials (like metals) probed by Dynamic Mechanical Analysis (DMA). More complex models can be analyzed in order to reproduce the behavior of other materials, or higher frequencies measurements. Their implication is essentially to generate a frequency-dependent Young's modulus $E(f)$ and friction term proportional to the rate of change of the stress tensor, which can thus be interpreted as an effective viscosity $\eta(f)$. 
As for the elastic constants, when introducing viscosity in a fluid one has to define (in the simplest case) \textit{two} constants: $\eta$ and  $\zeta$ the second viscosity\cite{landaufluids}. For the sake of simplicity, in our case we will use the same parametrization as for the elastic properties and we introduce a damping modulus $E_p$ with a damping Poisson's ratio $\nu_p$ (which are both also functions of frequency). 
As a consequence, when solving for a harmonic motion using complex forms, linear friction is \textit{equivalent to replacing} Young's modulus by a complex Young's modulus\cite{zenerbook} $E \rightarrow E + i E_2$: the linear mechanical response is de-phased from the excitation. 
This is a fairly simple writing which is widely used\cite{prl_unterrheitmeier,prl_regal}, but the price to pay is that all details of the friction model are hidden within the constant $E_2(f)$. 
Especially, its frequency-dependence (from first principles $E_2$ is function of $f$) is an important ingredient for the understanding of microscopic processes at stake.
The complete mathematical analysis can be found in Supplementary Information.

\vspace{0.2cm}
\begin{figure}
\includegraphics[width=\linewidth]{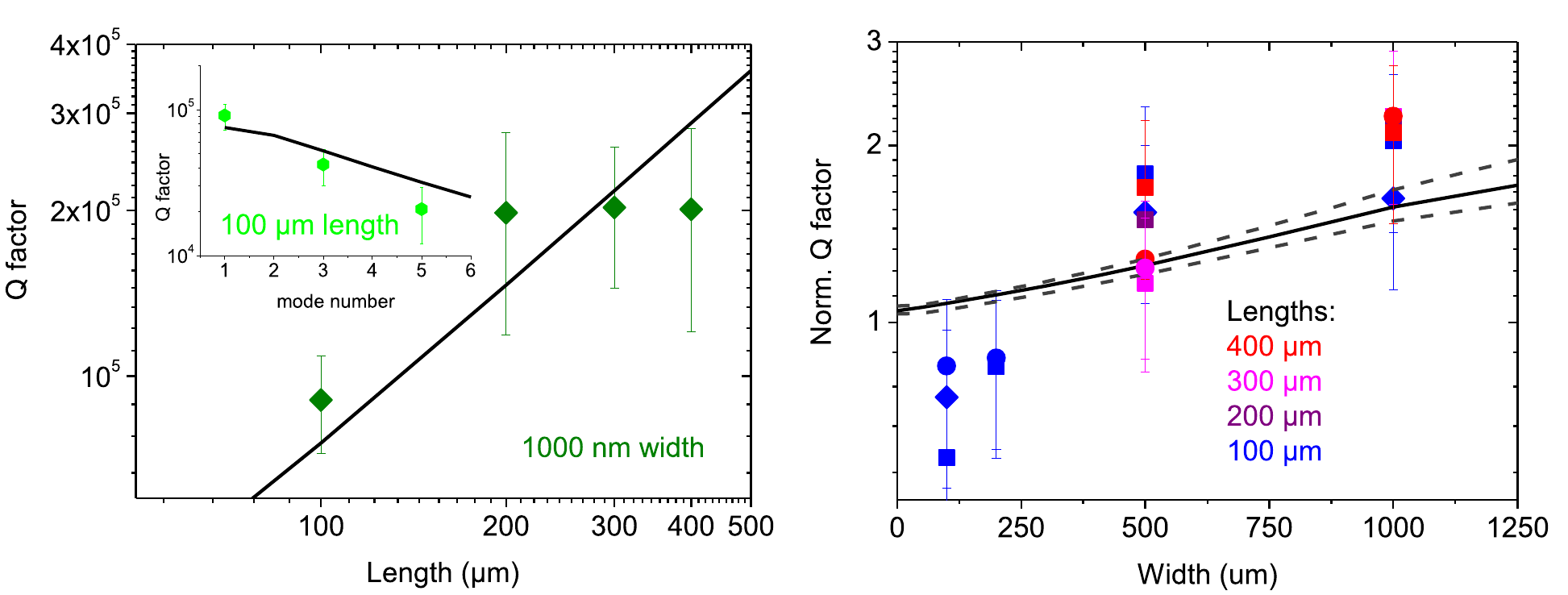}
\caption{\label{figure3} Left, main: 
dependence of quality factor on 
length $L$ for the first flexure $n=1$ (all devices having $w=1000~$nm). Left, inset: quality factor of 100$~\mu$m long device as a function of mode number $n$. Right: $Q$ normalized 
to the extrapolated value at $w \rightarrow 0$ for mode $n=1$ of all devices, as a function of width $w$. The lines correspond to Eqs. (\ref{fitQ},\ref{Dampfact},\ref{Qfact}) using a simple ansatz for the clamp parameter $\alpha$, see text.  }
\end{figure}
 
The quality factor $Q$ writes, for any mode $n$:
\begin{equation}
Q(n,a,\alpha) = Q_{n,0} \, \frac{P_f(n,a,\alpha)^2}{P_Q(n,a,\alpha)} , \label{fitQ}
\end{equation}
with $Q_{n,0} = f_{n,0}^2 / (f_{n,0} \Delta f_{n,0})$ the usual $n^{th}$ mode quality factor with:
\begin{equation}
f_{n,0} \Delta f_{n,0} = \frac{n^2}{4 \sqrt{3} } \, \frac{e}{L^3} \left( \frac{E_2}{\rho \sqrt{E/\sigma} } \right) . \label{Dampfact}
\end{equation}
Eq. (\ref{Dampfact}) has a single material-dependent fit parameter: $E_2$ ($\rho$, $E$ and $\sigma$ being known from resonance frequency fits). 
The $P_Q$ function is obtained as:
\begin{eqnarray}
P_Q(n,a,\alpha) & = & 1- \frac{6}{\alpha} + \frac{24 - n^2 \pi^2}{\alpha^2} \nonumber \\
& & \!\!\!\!\!\!\!\!\!\!\!\!\!\!\!\!\!\!\!\!\!\!\!\!\!\!\!\!\!\!\!\!\!\!\!\!\!\!\!\!\! + \, a \left(6+ \frac{1}{2} n^2 \pi^2 - \frac{48+6 n^2 \pi^2}{\alpha} +\frac{144-8 n^2 \pi^2}{\alpha^2 } \right) \! . \label{Qfact}
\end{eqnarray}
For ideal clamping ($\alpha \rightarrow \infty$), it is found experimentally that $Q$ scales linearly with $L$\cite{prl_unterrheitmeier}, and falls with increasing thickness (roughly as $1/e$ in Ref\cite{schmidPRB2011}.) and mode number $n$\cite{prl_unterrheitmeier}.
This has been fit with $E_2$ essentially frequency-independent\cite{prl_unterrheitmeier,prl_regal}; we reproduce this result in Supplementary Information by fitting the numerical solutions from Ref\cite{prl_unterrheitmeier}. Note the difference between Eqs. (\ref{fitQ},\ref{Dampfact},\ref{Qfact}) and the expressions derived for membranes\cite{prl_regal, kippNanoLett}.

Our data and theory are presented in Fig. \ref{figure3}. 
On the left panel, we show the conventional dependencies in $L$ and $n$ (the line is calculated with a fit $E_2$ value, see Supplementary Information). On the right panel, we demonstrate the soft clamping generated by our simple design: $Q$ grows substantially with increasing width $w$. The dashed upper
curve is calculated for 100$~\mu$m length, the lower one for 400$~\mu$m and the full line for 200$~\mu$m length.
The clamp parameter $\alpha$ used here is \textit{the same one} as for Fig. \ref{figure2}. 
The agreement is relatively good, but the remarkable result is that we predict theoretically the right tendency: the clamp degree of freedom reduces the bending at the anchor, which increases the $Q$. Introducing a boundary torsion spring $\gamma_l=\gamma_r=\gamma$ increases even further the effect; however, fitted values 
demonstrate that one requires a higher order expansion than the order 2 in torque for the quantitative result to be valid. We therefore preferred to keep this aspect outside of the present discussion.

\section{Conclusion}

We report on experiments performed on flexural nano-mechanical doubly-clamped beams, subject to an axial force load.
We demonstrate that their suspended anchoring points act as an ``easy soft clamping'': it is responsible for both a downward resonance frequency shift and an increase in quality factor.
We present an analytic theory that fits the data, based on simple boundary conditions: forces and torques undergone by the beam's ends, and due to the suspended clamp.
It turns out that the effective clamp spring constants $k_{l,r}$ are the dominant ingredient (with $l, r$ standing for left and right). 
For symmetric devices with similar clamps, $k_{l,r}\approx k$ is found to be $k \propto f/w$, with $f$ frequency and $w$ beam width.
For the frequency shift, following Ref\cite{VdZ_freq}. this can be recast in an effective lengthening $L \rightarrow L + \Delta L$ with $\Delta L \propto w/f$ at lowest order (and clamp-parameter dependent). For the quality factor, the fit is reasonably good but could be improved with a torque spring parameter $\Gamma_{l,r}$. To do so, the theory would need to be improved; likewise, it could be extended to the case of beams with no (or very little) axial stress, and even to cantilevers. 
The problem addressed here is extremely 
widespread, 
and beams with a ``natural'' clamp undercut (i.e. due to the fabrication process) are extensively used.
For this reason, we believe that our work is very 
relevant to nano-mechanical design, for both defining precisely resonance frequencies (mandatory when multiplexing is at hand) and quality factors.

\section{Author Information}

\textbf{Corresponding Author} \\
$^*$: Eddy Collin, eddy.collin@neel.cnrs.fr. \\

\textbf{Notes} \\
The authors declare no competing financial interest.\\
Data Availability: the data that support the findings of this study are openly available in {\bf Cloud N\'eel} at 
https://cloud.neel.cnrs.fr/index.php/s/CnnYPKn8XHYZgXa, reference number \cite{cloud}.
\begin{acknowledgement}

We acknowledge the use of the N\'eel facility \textit{Nanofab} for the devices fabrication, and the N\'eel \textit{Cryogenics} facility for help with the low temperature setup. 
The authors acknowledge support from the ERC CoG grant ULT-NEMS No. 647917, and ERC StG grant UNIGLASS No. 714692. 
The research leading to these results has received funding from the European Union's Horizon 2020 Research and Innovation Programme, under grant agreement No. 824109, the European Microkelvin Platform (EMP).



\end{acknowledgement}

\begin{suppinfo}

Supplementary Information document: provides details about the samples, and the complete description of the mathematical modeling.
The analytic theory (derived using Mathematica\textsuperscript{\textregistered}) is compared to published theoretical results with ideal clamping conditions.
Numerical finite element simulations (COMSOL\textsuperscript{\textregistered}) are also presented, reproducing qualitatively the data.


\newpage
\section{Modal expansion}

We start by presenting the theoretical modeling. We consider the case of thin-and-long beams, i.e. $w \ll L$ and $e \ll L$ ($w$ width, $e$ thickness and $L$ length).

\begin{figure}[H]
    \centering
    \includegraphics[width=8.cm]{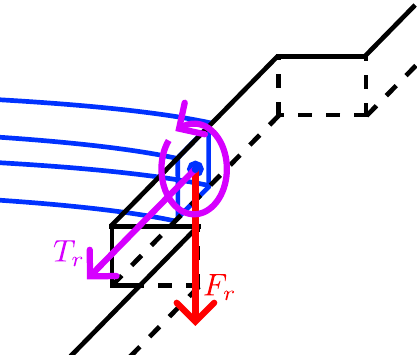}
    \caption{Schematic of (right) clamp with force $F_r$ and torque $T_r$ that the suspended part (black) exerts on the beam (in blue, see text).}
    \label{fig:my_clamp}
\end{figure}

From the well-known Euler-Bernoulli equation (including both bending and stress terms), we describe the flexure $f(z,t)$:
\begin{equation}
        E I_z \frac{\partial^4 f(z,t)}{\partial z^4} - S \frac{\partial^2 f(z,t)}{\partial z^2} = - \rho A \frac{\partial^2 f(z,t)}{\partial t^2}
    \label{eq_eulerbernoulli}
\end{equation}
where $A=we$ - cross-section area, $I_z=\frac{1}{12}we^3$ - second moment of area, $E$ - material's Young's modulus, $S$ - inside tensile $(S>0)$ force of the material, $\rho$ - material density.

We are looking for solution as $f(z,t) = \psi_n(z)x_n(t)$, where $x_n(t)=$ \linebreak 
$x_0\cos{(\omega_nt + \varphi)}$ is the temporal harmonic part (out-of-plane motion amplitude) and $\psi_n(z)$ - $n$'s mode shape. $\omega_n$ is the mode resonance frequency.

On each side, we introduce a force $F_{l,r}$ and torque $T_{l,r}$, see Fig. \ref{fig:my_clamp}.
The relaxed boundary conditions with added spring constant $k_{l,r}$ and inertia $m_{l,r}$ write, for the left $l$ and right $r$ clamps (and similarly for the torques):\\

\textbf{Forces:}
\begin{align}
  F_l=  +E I_z \frac{\partial^3 f(z=0,t)}{\partial z^3} - S \frac{\partial f(z=0,t)}{\partial z} &= - k_{l}f(z=0,t) + m_{l} \frac{\partial^2 f(z=0,t)}{\partial t^2}\\
  F_r=  -E I_z \frac{\partial^3 f(z=L,t)}{\partial z^3} + S \frac{\partial f(z=L,t)}{\partial z} &= - k_{r}f(z=L,t) + m_{r} \frac{\partial^2 f(z=L,t)}{\partial t^2}
\end{align}

\textbf{Torques:}
\begin{align}
   T_l= -E I_z \frac{\partial^2 f(z=0,t)}{\partial z^2}  &= - \Gamma_{l}\frac{\partial f(z=0,t)}{\partial z} + \mathcal{M}_{l} \frac{\partial^3 f(z=0,t)}{\partial t^2 \partial z}\\
   T_r= +E I_z \frac{\partial^2 f(z=L,t)}{\partial z^2} &= - \Gamma_{r}\frac{\partial f(z=L,t)}{\partial z} + \mathcal{M}_{r} \frac{\partial^3 f(z=L,t)}{\partial t^2\partial z}
\end{align}
Note signs in forces and torques definition.

The general solution writes as:
\begin{align}
    \psi_n(z) &= C_{n,1}\sin{\left(k_{n+}\frac{z}{L}\right)} +C_{n,2}\cos{\left(k_{n+}\frac{z}{L}\right)} \nonumber \\
    & + C_{n,3}\sinh{\left(k_{n-}\frac{z}{L}\right)} +C_{n,4}\cosh{\left(k_{n-}\frac{z}{L}\right)}
\end{align}
Considering a shape normalised to $1$: $max[\psi(z)]=1$ at $z=z_{max}$ (we chose $0<z_{max} \leq L/2$ with no loss of generality).

Now we expand in the high-stress limit, which means having a small parameter $a=\sqrt{\frac{E I_Z}{SL^2}}\ll 1$:
\begin{equation}
    \omega_n = \frac{k_n(a)}{L}\sqrt{\frac{\sigma}{\rho}}
\end{equation}
where the stress is defined as $\sigma = S/A$.

$k_{n-}$ and $k_{n+}$ are deduced  from $k_n(a)$ and they all can be written as Taylor expansions:
\begin{align}
    k_n(a) &= k_n(0)+k_n'(0)\cdot a + (1/2)k_n''(0)\cdot a^2 + \dots\\
    k_{n+}(a) &= k_n(0)+k_n'(0)\cdot a + (1/2)[k_n''(0) -k_n^3(0)]\cdot a^2 + \dots\\
    k_{n-}(a) &= 1/a + (1/2)k_n^2(0)\cdot a + [k_n(0) k_n'(0) ]\cdot a^2 + \dots
\end{align}
Note the 
$1/a$ in the last Eq. which diverges for small $a$. In the $\cosh$ and $\sinh$ functions, it should be treated with care: we shall neglect $\exp{\left(-X\right)}$ terms with $X \propto 1/a$ (what we call exponential approximation), and after take series expansions in $a$ (to lowest order up to $a^2$). Note that the first order terms goes as $a$, while the low-stress equivalent expansion leads to small term $1/a^2 = S L^2/(E I_z)$. An equivalent modeling can be performed in this limit, but is outside of the scope of the paper.

Solving the problem produces the mode shape with the definition of constants $C_{n,i}$ and $z_{max}$, with $k_n(0), k_n'(0), k_n''(0)$. This is performed using an extensive Mathematica\textsuperscript{\textregistered} code.

\begin{figure}[H]
    \centering
    \includegraphics[width=11.cm]{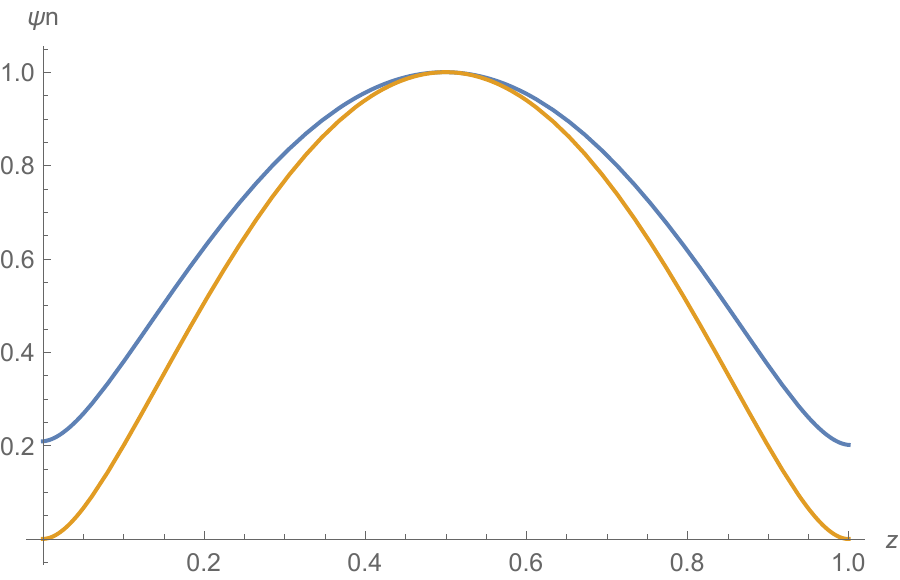}
    \caption{Mode shape with ideal clamping in orange, and with $\alpha_l=\alpha_r=15.$, $\gamma_l=\gamma_r=1.$ in blue (see text; in this case, $z_{max}=L/2$ in both cases). For this graphics,  $a=0.05$ and the $a$ expansion has been pushed to order 4 for the ideal clamping terms.}
    \label{fig:my_shape}
\end{figure}

Regrouping restoring forces and inertial components, we fit with for each side two fit parameters:
\begin{eqnarray}
k_{r,l} & \rightarrow & k_{r,l} + m_{r,l}\, \omega^2  \\
\Gamma_{r,l} & \rightarrow & \Gamma_{r,l} + {\cal M}_{r,l}\, \omega^2  
\end{eqnarray}
without losses of generality. Introducing dimensionless parameters, $k_{r,l}=\frac{S}{L} \alpha_{r,l}$ and $\Gamma_{r,l} = (S L)\, \gamma_{r,l}$. The solutions in the following Sections are given with an expansion at order 2 in $1/\alpha_{r,l} \ll 1$ and $1/\gamma_{r,l} \ll 1$. As an example, we show in Fig. \ref{fig:my_shape}
the ideal shape with perfect clamping ($\alpha_{l,r}$ and $\gamma_{l,r}$ being $\infty$), and the one obtained for imperfect clamps (see caption).
The shape is not zero anymore at both ends $z=0,L$, and (more subtle) the angle $\partial f/\partial z $ is not either.

\section{Stored energy}

Defining energies from forces (and force densities):
\begin{align}
    F_{flex} = +E  I_z \frac{\partial^3 f}{\partial z^3} \longrightarrow
    \frac{\partial F}{\partial z} = +E  I_z \frac{\partial^4 f}{\partial z^4} \longrightarrow 
    \mathcal{E}_{flex} &= \frac{1}{2}E I_z\int_{0}^{L} \left(\frac{\partial^4 f}{\partial z^4}f \right) dz \\
    F_{tens} = -S \frac{\partial f}{\partial z} \longrightarrow
    \frac{\partial F}{\partial z} = -S \frac{\partial^2 f}{\partial z^2} \longrightarrow 
    \mathcal{E}_{tensile} &= \frac{1}{2}S\int_{0}^{L} \left(-\frac{\partial^2 f}{\partial z^2} f \right) dz\\
    \rho A \frac{\partial^2f}{\partial t^2} \longrightarrow
    \mathcal{E}_{kin} &= \frac{1}{2}\rho A \int_{0}^{L} \left(\frac{\partial f}{\partial t}\right)^2 dz
\end{align}

Now from integration by parts for flexural energy:
\begin{align}
    \left( \frac{\partial^2 f}{\partial z^2} \cdot \frac{\partial f}{\partial z}\right)' &= \frac{\partial^3f}{\partial z^3}\cdot \frac{\partial f}{\partial z} +\left(\frac{\partial^2 f}{\partial z^2} \right)^2\\
    \left( \frac{\partial^3 f}{\partial z^3} \cdot f\right)' &= \frac{\partial^4 f}{\partial z^4}\cdot f + \frac{\partial^3 f}{\partial z^3}\cdot \frac{\partial f}{\partial z}
\end{align}

From which follows:
\begin{align}
    \frac{\partial^4 f}{\partial z^4}\cdot f = \left( \frac{\partial^3 f}{\partial z^3} \cdot f - \frac{\partial^2 f}{\partial z^2} \cdot \frac{\partial f}{\partial z} \right)' + \left(\frac{\partial^2 f}{\partial z^2} \right)^2 \\
    \int_0^L \frac{\partial^4 f}{\partial z^4}\cdot f \ dz = \underbrace{\left[  \frac{\partial^3 f}{\partial z^3} \cdot f - \frac{\partial^2 f}{\partial z^2} \cdot \frac{\partial f}{\partial z}  \right]_0^L}_{\mathrm{boundary \ term}} + \underbrace{\int_0^L \left(\frac{\partial^2 f}{\partial z^2} \right)^2 \ dz}_{bulk \ term}
\end{align}

Same for tensile energy:
\begin{align}
    \left(\frac{\partial f}{\partial z} \cdot f \right)' &= \frac{\partial^2 f}{\partial z^2} \cdot f + \left( \frac{\partial f}{\partial z} \right)^2\\
    \frac{\partial^2f}{\partial z^2} \cdot f &= \left( \frac{\partial f}{\partial z} \cdot f \right)'  +\left[ -\left(\frac{\partial f}{\partial z}\right)^2 \right]\\
    \int_0^L\left( -\frac{\partial^2f}{\partial z^2} \cdot f \right) \ dz &= \underbrace{\left[ -\frac{\partial f}{\partial z} \cdot f \right]_0^L}_{boundary \ term} +  \underbrace{\int_0^L \left( \frac{\partial f}{\partial z}  \right)^2 \ dz}_{bulk \ term}
\end{align}

We therefore can define a bulk term and a boundary term for these energies. The boundary term should be $=0$ for an ideal clamp, i.e. $f=0$ and $\partial f/\partial z = 0$ at $z=0,L$. But here, it is not the case; we define for each side:
\begin{align}
    \mathcal{E}_{flex,bound} = &+\frac{E I_z}{2} \frac{\partial^3f(z=0,t)}{\partial z^3} f(z = 0,t)\\ &- \frac{E I_z}{2} \frac{\partial^2f(z=0,t)}{\partial z^2} \frac{\partial f(z=0,t)}{\partial z}\nonumber\\
    \mathcal{E}_{tens,bound} = &- \frac{S}{2} \frac{\partial f(z=0,t)}{\partial z} f(z=0,t)
\end{align}
and same for $z=L$, but with reversed signs. One has to pay attention to signs with the definitions and the integration by parts procedure.

Then we define effective energies for the mode, as a function of the mode amplitude $x_n$:
\begin{align}
    \mathcal{E}_{flex} +\mathcal{E}_{tens} &= \frac{1}{2} k_n x_n^2(t)\\
    \mathcal{E}_{kin} &= \frac{1}{2} m_n \Dot{x}_n^2(t) 
\end{align}
Which include in $\mathcal{E}_{flex}$ and $\mathcal{E}_{tens}$ the bulk and boundary terms.
One should obviously recover $\omega_n^2 = k_n/m_n$, where $k_n$ is the mode's spring constant and $m_n$ the mode's effective mass per definition, computed from the mode shape $\psi_n$. This has been explicitly checked in the Mathematica\textsuperscript{\textregistered} code. In the literature, this is usually not mentioned.

\begin{figure}[H]
    \centering
    \includegraphics[width=11.cm]{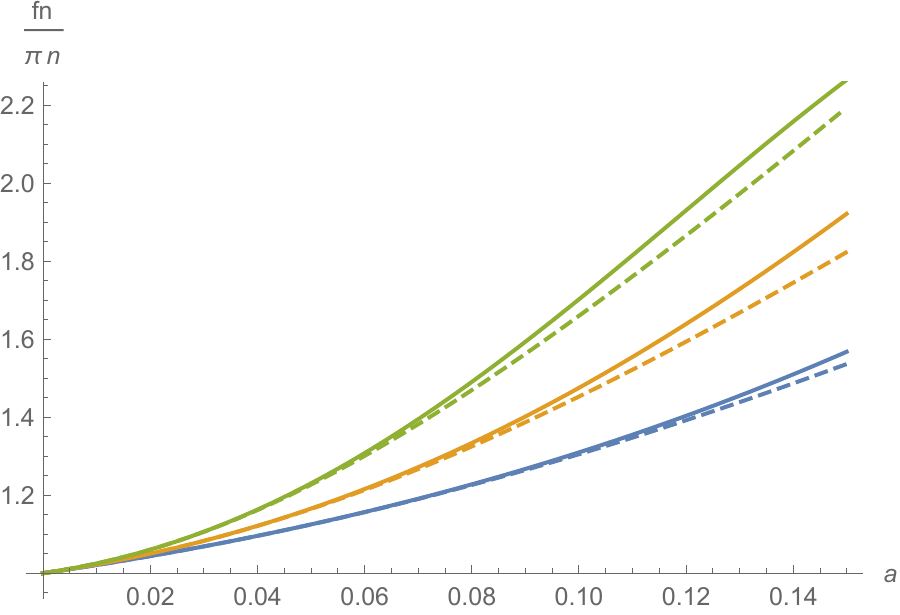}
    \caption{Comparison in the ideal clamp limit of frequency expansion (full lines) to the exact numerical result by [Bokaian, J. Sound and Vibr. {\bf 142}, 481 (1990)] (dashed lines). Blue $n=1$, orange $n=2$ and green $n=3$, as a function of $a$ (see text).}
    \label{fig:my_bokaian}
\end{figure}

In the first place, the modeling allows us to compute the resonance frequencies.
The full expression at second order in all parameters reads, for the deviation function $P_f$ introduced in the core of the paper:

\begin{eqnarray}
P_f(n,a,\alpha) & = & 1- \left(\frac{1}{\alpha_l}+\frac{1}{\alpha_r}\right) + \left(\frac{1}{\alpha_l}+\frac{1}{\alpha_r}\right)^2 \nonumber \\
&& \!\!\!\!\!\!\!\!\!\!\!\!\!\!\!\!\!\!\!\!\!\!\!\!\!\!\!\!\!\!\!\!\!\!\!\!\!\!\!\!\!\!\!\!\!\!\! + \, a \left[2 - 4\left( \frac{1}{\alpha_l}+\frac{1}{\alpha_r}\right) + \left( \frac{6- n^2 \pi^2}{\alpha_l^2}+\frac{6- n^2 \pi^2}{\alpha_r^2}+\frac{12}{\alpha_l \alpha_r}\right) \right]  \nonumber \\
&& \!\!\!\!\!\!\!\!\!\!\!\!\!\!\!\!\!\!\!\!\!\!\!\!\!\!\!\!\!\!\!\!\!\!\!\!\!\!\!\!\!\!\!\!\!\!\! + \frac{1}{2} \,a^2 \left[ 8+ n^2 \pi^2 -(24+5 n^2 \pi^2) \left( \frac{1}{\alpha_l}+\frac{1}{\alpha_r}\right)  \right. \nonumber \\
&&\!\!\!\!\!\!\!\!\!\!\!\!\!\!\!\!\!\!\!\!\!\!\!\!\!\!\!\!\!\!\!\!\!\!\!\!\!\!\!\!\!\!\!\!\!\!\! \left. + \left( \frac{48- 2n^2 \pi^2}{\alpha_l^2}+\frac{48- 2n^2 \pi^2}{\alpha_r^2}+\frac{96+28 n^2 \pi^2}{\alpha_l \alpha_r}\right) \right. \nonumber \\
&&\!\!\!\!\!\!\!\!\!\!\!\!\!\!\!\!\!\!\!\!\!\!\!\!\!\!\!\!\!\!\!\!\!\!\!\!\!\!\!\!\!\!\!\!\!\!\! \left. -2\left(\frac{1}{\gamma_l}+\frac{1}{\gamma_r}\right)+4\left(\frac{1}{\alpha_l}+\frac{1}{\alpha_r}\right)\left(\frac{1}{\gamma_l}+\frac{1}{\gamma_r}\right) \right]  \label{freq}
\end{eqnarray}

Note the symmetry $l \leftrightarrow r$ in this expression. A simplified version of it is in the main paper, Eq. (3).
In the case of an ideal clamp (effective normalised spring constants $\rightarrow \infty$), 
we compare the expansion to the exact numerical result from  [Bokaian, J. Sound and Vibr. {\bf 142}, 481 (1990)] in Fig.
\ref{fig:my_bokaian}.
For the plot, we normalise the frequency to $n \pi$, and take $1/L \sqrt{\sigma/\rho}=1$.
Our computed frequencies match the exact result for typically $a<0.1$ in the case of ideal clamps, with $n=1$. And it gets worse for higher modes $n$, which can be compensated for by taking higher order terms in the $a$-expansion. For the sake of making a good plot, this is why we used a higher expansion in Fig. \ref{fig:my_shape} (up to fourth order).

\section{Dissipated energy}

We shall use the same approach as the one proposed in [Quirin P. Unterreithmeier, Thomas Faust, and Jörg P. Kotthaus PRL \textbf{105} 027205 (2010)]. But we formalise it from the continuum mechanics approach as presented in A.N. Cleland, {\it Foundations of Nanomechanics}, Springer 2003. This implies some minor modification:
\begin{itemize}
    \item we make it more generic and exact by starting from stress and strain; Poisson's ratios will appear in the expression,
    \item we start from a proper friction force definition, as in  Zener's model,  generating a friction force proportional to the rate of change of the stress. This is what is behind the commonly used complex Young's modulus.
\end{itemize}

Let us start with the strain 6-component vector for motion of the beam with mode shape $f(z,t)$ and distortion in the  $\Vec{x}$ direction (see A.N. Cleland  2003):
\begin{equation}
    \varepsilon = (\nu x f'', \nu x f'',-xf'',0,0,\nu x f'') .
\end{equation}
This expression essentially assumes that planes orthogonal to the neutral axis remain orthogonal to the displaced neutral axis (Euler-Bernoulli approximation), while it guarantees no lateral stresses (see $\sigma_{el}$ below).

Then the linear response for the stress is:
\begin{equation}
    \sigma_{el} = (0,0,-E xf'',0,0,\frac{E  \,\nu}{2(1+\nu)}x f'')
\end{equation}
Where we have both Young's $E $ and Poisson's $\nu$ (elasticity theory). We assume linearity to apply, such that this stress adds up to the in-built load $\sigma$ (purely along $\vec{z}$).

The same linear hypothesis can be made for the local friction stress, which at the macroscopic scale will produce the friction force (prop. to $\dot{x_n}$). But in order to match Zener's low frequency limit, and reproduce a standard viscous friction mechanism, this friction component shall be proportional to the time rate of change of strain $\dot{\varepsilon}$.
For the friction component we thus similarly introduce:
\begin{align}
    \sigma_{fr} =E_p \cdot \Bigg(&\frac{ \nu-\nu_p}{(1+\nu_p)(1-2\nu_p)},\frac{ \nu-\nu_p}{(1+\nu_p)(1-2\nu_p)}, \\ &-\frac{ 1-\nu_p-2 \nu \nu_p}{(1+\nu_p)(1-2\nu_p)} ,0,0,\frac{ \nu}{2(1+\nu_p)}\Bigg) \cdot x \dot{f}''\nonumber
\end{align}
with two parameters describing fricion, $E_p$ and $\nu_p$ (similar parametrization of linear response as for stress-strain). 
Note that the origin of these terms is outside of the scope of the modeling; but obviously $E_p(\omega)$ and $\nu_p(\omega)$ depend on frequency, since they originate in microscopic mechanisms which should depend on $\omega$ (see comment below).

The power density lost in friction is then:
\begin{align}
    \mathcal{P} = \sigma_{fr}\cdot \dot{\varepsilon} &= E_p \left( 1 + \frac{\nu^2(5-2 \nu_p)-8 \nu \nu_p +4 \nu_p^2 }{2(1+\nu_p)(1-2\nu_p)} \right) x^2 \left( \dot{f}'' \right)^2 \\ &= 
    E_p \left( 1 + \underbrace{\frac{\nu^2(5-2 \nu_p)-8 \nu \nu_p +4 \nu_p^2 }{2(1+\nu_p)(1-2\nu_p)}}_{\mathrm{small \ parameter \ }o(\nu,\nu_p)} \right) \omega^2 x^2 \left( \frac{\partial^2 \psi_n(z)}{\partial z^2} \right)^2 x_n^2(t)\nonumber
\end{align}

Now integration over cross-section $x^2\rightarrow I_z$ and length $\int_0^L$:
\begin{equation}
    \iiint  \mathcal{P} \ dxdydz = E_p\left[ 1 + o(\nu,\nu_p) \right] \, \omega^2 \,I_z \underbrace{\int_0^L\left( \frac{\partial^2\psi_n}{\partial z^2}\right)^2 dz}_{\mathrm{same \ integral \ as \ in \ bending \ energy}} \  x_n^2(t) 
\end{equation}

In terms of mode definition it can be rewritten introducing the friction force:
\begin{equation}
    \iiint \mathcal{P} \ dxdydz = \underbrace{\Lambda_n \dot{x}_n}_{\mathrm{eff. \ friction \ force}} \cdot \dot{x}_n = \Lambda_n \omega^2 x^2_n(t)
\end{equation}

By definition damping parameter $\Delta \omega_n$ is (and $Q=\omega_n/\Delta \omega_n$):
\begin{equation}
    \Delta \omega_n = \frac{\Lambda_n}{m_n}
\end{equation}

Therefore we can identify:
\begin{equation}
    \Lambda_n = \underbrace{E_p \left[ 1 + o(\nu,\nu_p) \right]}_{\mathrm{fit. \ parameter}}\, I_z \int_0^L \left( \frac{\partial^2 \psi_n}{\partial z^2} \right)^2 dz
\end{equation}

There is one formal difference here with the [Quirin P. Unterreithmeier, Thomas Faust, and Jörg P. Kotthaus PRL \textbf{105} 027205 (2010)] modeling, which uses a compex Young's modulus with imaginary part $E_2$:
their modeling is equivalent to the above one with the identity $E_p \left[ 1 + o(\nu,\nu_p) \right] = E_2/\omega$.
As a result, with $\omega \approx \omega_n$ for a high-$Q$ resonance we have:
\begin{equation}
    \omega_n \Delta \omega_n =  \frac{e^2 E_2}{L^4 \rho}\frac{1}{12} \frac{ \int_0^1 \left( \frac{\partial^2 \psi_n[\tilde{z}]}{\partial \tilde{z}^2} \right)^2 d\tilde{z} }{ \int_0^1 \left(\psi_n[\tilde{z}] \right)^2 d\tilde{z}}
\end{equation}
with $\tilde{z}=z/L$.
The shape factor defined from $\psi_n$ (the ratio of integrals on the right) tends to $2 n^2 \pi^2/a$ at lowest order. This leads to the result Eq. (6) of the main paper, written in Hz (here, we have all expressions in Rad/s).
From this, one defines the quality factor $Q$, and then the function $P_Q$:

\begin{eqnarray}
P_Q(n,a,\alpha) & = & 1-3\left( \frac{1}{\alpha_l}+ \frac{1}{\alpha_r}\right) + \left(\frac{6 -\frac{1}{2} n^2 \pi^2}{\alpha_l^2}+ \frac{6 -\frac{1}{2} n^2 \pi^2}{\alpha_r^2}+ \frac{12}{\alpha_l \alpha_r} \right) \nonumber \\
& & \!\!\!\!\!\!\!\!\!\!\!\!\!\!\!\!\!\!\!\!\!\!\!\!\!\!\!\!\!\!\!\!\!\!\!\!\!\!\!\!\! + \, a \left[ 6+ \frac{ n^2 \pi^2}{2} - (24+3 n^2 \pi^2) \left( \frac{1}{\alpha_l}+\frac{1}{\alpha_r} \right) +\left( \frac{60}{\alpha_l^2 }+ \frac{60}{\alpha_r^2 }+ \frac{24-8 n^2 \pi^2}{\alpha_l \alpha_r } \right) \right. \nonumber \\
&& \left. \!\!\!\!\!\!\!\!\!\!\!\!\!\!\!\!\!\!\!\!\!\!\!\!\!\!\!\!\!\!\!\!\!\!\!\!\!\!\!\!\!-\left( \frac{1}{\gamma_l}+\frac{1}{\gamma_r} \right) +3 \left( \frac{1}{\alpha_l}+\frac{1}{\alpha_r} \right) \left( \frac{1}{\gamma_l}+\frac{1}{\gamma_r} \right) \right]  \label{Qfact}
\end{eqnarray}
which represents all deviations from the basic expression (at lowest order in $a$; note the $l \leftrightarrow r$ symmetry). A simplified version is in the main paper, Eq. (7).
In the limit where the clamp spring constants $\alpha_{l,r}, \gamma_{l,r}$ tend to $\infty$, one recovers the usual tendencies: a $Q$ factor that grows as $L$, and decreases with mode number $n$. We demonstrate this by comparing our formulas to the numerical results of [Quirin P. Unterreithmeier, Thomas Faust, and Jörg P. Kotthaus PRL \textbf{105} 027205 (2010)]. This is done in Fig. \ref{fig:my_unterrheitmeier}, see Caption for the used numerical values (we take here a  $E_2$ independent of $\omega$). The agreement is very good; we believe deviations are due to the finite order of the expansion in $a$.

Further fits of our own data (including the clamp dependence) are discussed in the next Section.
\begin{figure}[H]
    \centering
    \includegraphics[width=12.cm]{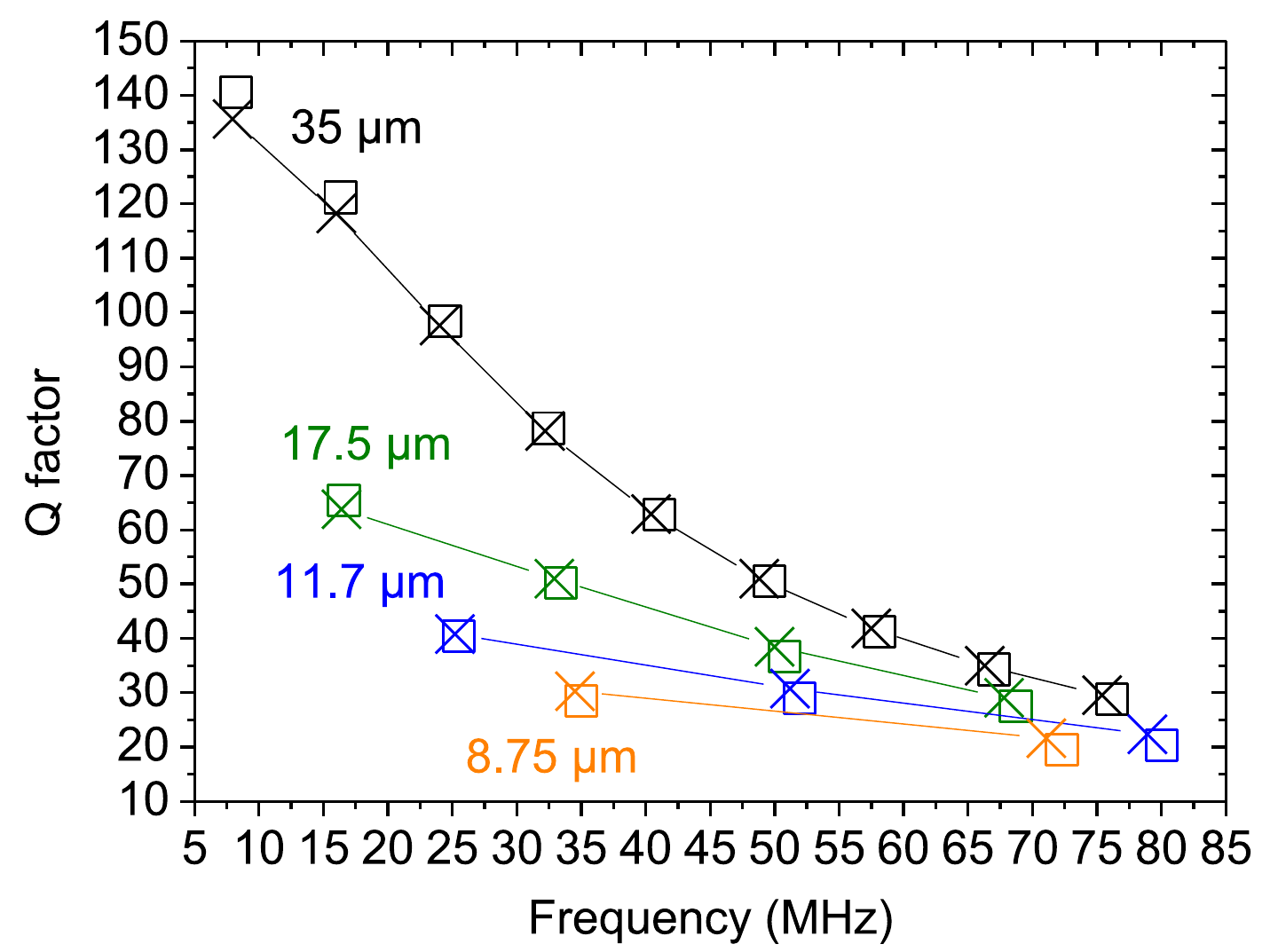}
    \caption{Numerical calculation of $Q,f_0$ from [Quirin P. Unterreithmeier, Thomas Faust, and Jörg P. Kotthaus PRL \textbf{105} 027205 (2010)] (open squares, for different beam lengths $L$) compared to our analytic solution (crosses), in the case of ideal clamp. Parameters chosen: $\sigma=0.83~$GPa, $E=160~$GPa, $\rho=2800~$kg/m$^3$, $E_2=48~$MPa, $e=100~$nm, $w=200~$nm.}
    \label{fig:my_unterrheitmeier}
\end{figure}

\section{Sample Characteristics and clamp fit}

The sample fabrication and main characteristics are given in [Golokolenov et al. JLTP 2022].
The measurements have been performed on two chips, on which eight sets of beams (2 groups with the 4 different lengths) were present. Not all of them have been measured (for lack of time), but a good fraction of them has been characterised. Out of all the data acquired, only few data points have been excluded, because they seemed to be out-of-statistics (frequency off by about 20$~\%$ max., or $Q$ factor very low and almost independent of parameters $w,n$). We believe that this comes from the reproducibility of the fabrication, which remains our limiting parameter.

All data in the core of the paper has been fit with a single set of parameters, including a single (symmetric) clamp ansatz. We chose:
\begin{itemize}
\item $\sigma = 0.17~$GPa (low-stress sample),
\item $\rho=2800~$kg$/$m$^3$,
\item $E=160~$GPa,
\item $E_2=180~$MPa,
\end{itemize}
that describe the properties of the bilayer structure. Note that our $E_2$ is larger than the one of [Quirin P. Unterreithmeier, Thomas Faust, and Jörg P. Kotthaus PRL \textbf{105} 027205 (2010)] which had no metal layer. The clamp spring constant is found to be:
\begin{equation}
\alpha_l=\alpha_r=\alpha \approx 2500. \frac{n \pi}{w}, \label{alpha}
\end{equation}
in order to reproduce the properties measured (see discussion in main paper). We verify $\alpha \gg 1$ in the whole range studied. For the sake of completeness, we also included a torque spring, following the same ansatz. However, we realised that in order to influence the  fit, this one had to be particularly small, i.e. out of the validity range of the expansion. We therefore preferred not to discuss this point further in the paper; because it would require a more exact expansion to be included. Technically, the graphs of the paper are obtained with:
\begin{equation}
\gamma_l=\gamma_r=\gamma \approx 300. \frac{n \pi}{w},  \label{gamma}
\end{equation}
which has an almost invisible impact on fits (and becomes $\sim 1$ at worst).
This gives us a sort of estimate of the limit within which torques can be neglected.

\begin{table}[h!]
\begin{center}
\begin{tabular}{|c|c|c|c|}    \hline
Sample name    &  left clamp ($\mu$m) &  right clamp ($\mu$m) &  mean ($\mu$m) \\    \hline    \hline
 S2 100$~\mu$m assym. &  3.85        &  13.1        & 8.5         \\    \hline
 S2 100$~\mu$m sym.   &  7.7        &  11.55        & 9.6         \\    \hline
 S2 300$~\mu$m        &  10.        &  14.          & 12.         \\    \hline
 S2 400$~\mu$m        &  5.8        &  13.5         & 9.6         \\    \hline
 S6 100$~\mu$m        &  12.5        &  16.7        & 14.6         \\    \hline
 S6 200$~\mu$m        &  10.8        &  12.3        & 11.5         \\    \hline
 S6 300$~\mu$m        &  13.1        &  16.9        & 15.         \\    \hline
 S6 400$~\mu$m        &  11.3        &  16.2        & 13.7         \\    \hline
\end{tabular}
\caption{\label{table} Clamp suspension length $L_c$ statistics.}
\end{center}
\end{table}

Eqs. (\ref{alpha},\ref{gamma}) are characteristics of the clamp. Within our geometry, its thickness is constant, its composition is the same as for the beam, and we assume that its width $W$ does not matter either, since $w \ll W$ (at worst about 20$~\mu$m): for the distortion of the anchoring point, the width of the clamp is essentially infinite.
But the length $L_c$ of the suspended part should obviously matter; we give in Tab. \ref{table} the statistics of the clamp's suspended region, for all samples. On average, we have $L_c \approx 12~\mu$m, within about $\pm 50~\%$ scatter in the fabrication. We believe that this scatter could also contribute to the dispersion in the measured properties.
\begin{figure}[H]
    \centering
    \includegraphics[width=12.cm]{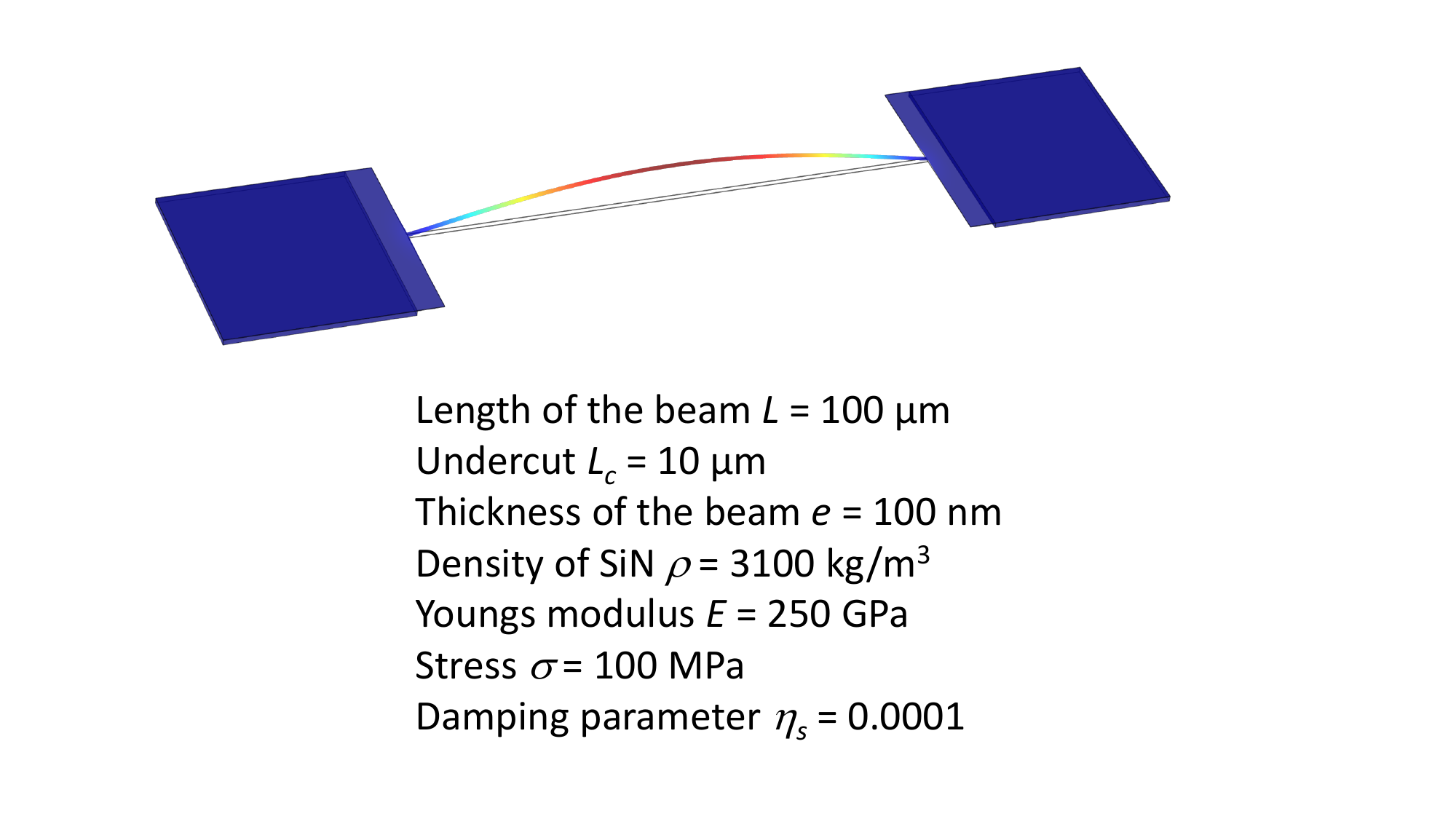}
    \caption{COMSOL\textsuperscript{\textregistered} simulation (first $n=1$ flexure) reproducing the "easy soft clamping" effect (see text). Parameters used for simulation given in Figure.}
    \label{fig:my_comsol}
\end{figure}

\section{Comparison to Numerical Simulations}

The measured properties can also be qualitatively reproduced by a numerical simulation, demonstrating the "easy soft clamping" effect. To demonstrate this, we perform a simple finite element study on COMSOL\textsuperscript{\textregistered}, as shown in Fig. \ref{fig:my_comsol}.
Parameters used given in the figure, except for width $w$ which is varied.

The calculated resonance frequencies $f_0$ and quality factors $Q$
are presented (in normalised from) in Fig. \ref{fig:my_fitsnums}, as a function of $w$.
The very same trend as in the experimental data is visible: the frequency drops with increasing $w$, while the quality factor grows. 
It can be fit by a second order polynomial in $w$, as expected from our analytic theory. The magnitude of the effect is also correct as compared to experiments, but is not in perfect quantitative agreement.
We believe that this is due to the exact, and rather precise, choice of parameters that one has to perform to match simulations on data.

\begin{figure}[H]
    \centering
    \includegraphics[width=12.cm]{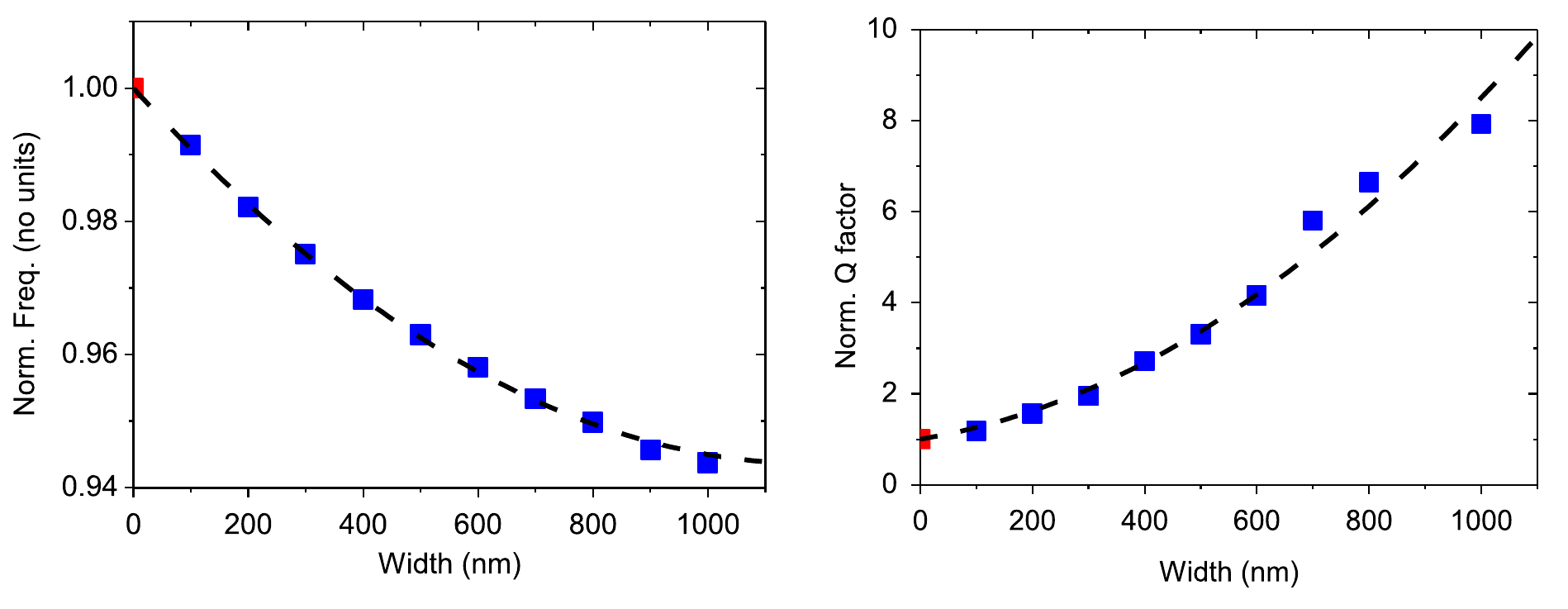}
    \caption{Normalised frequency (left) and quality factor (right) as a function of beam width. The red point is the $w=0$ extrapolated value, and the dashed lines simple polynomial (second order) fits (see text). }
    \label{fig:my_fitsnums}
\end{figure}

\end{suppinfo}

\bibliography{achemso-demo}

\providecommand{\latin}[1]{#1}
\makeatletter
\providecommand{\doi}
  {\begingroup\let\do\@makeother\dospecials
  \catcode`\{=1 \catcode`\}=2 \doi@aux}
\providecommand{\doi@aux}[1]{\endgroup\texttt{#1}}
\makeatother
\providecommand*\mcitethebibliography{\thebibliography}
\csname @ifundefined\endcsname{endmcitethebibliography}
  {\let\endmcitethebibliography\endthebibliography}{}
\begin{mcitethebibliography}{40}
\providecommand*\natexlab[1]{#1}
\providecommand*\mciteSetBstSublistMode[1]{}
\providecommand*\mciteSetBstMaxWidthForm[2]{}
\providecommand*\mciteBstWouldAddEndPuncttrue
  {\def\EndOfBibitem{\unskip.}}
\providecommand*\mciteBstWouldAddEndPunctfalse
  {\let\EndOfBibitem\relax}
\providecommand*\mciteSetBstMidEndSepPunct[3]{}
\providecommand*\mciteSetBstSublistLabelBeginEnd[3]{}
\providecommand*\EndOfBibitem{}
\mciteSetBstSublistMode{f}
\mciteSetBstMaxWidthForm{subitem}{(\alph{mcitesubitemcount})}
\mciteSetBstSublistLabelBeginEnd
  {\mcitemaxwidthsubitemform\space}
  {\relax}
  {\relax}

\bibitem[Sage \latin{et~al.}(2018)Sage, Sansa, Fostner, Defoort, Marc~G\'ely,
  Morel, Duraffourg, Roukes, Alava, Jourdan, Colinet, Masselon, Brenac, and
  Hentz]{roukes_mass}
Sage,~E.; Sansa,~M.; Fostner,~S.; Defoort,~M.; Marc~G\'ely,~A. K.~N.;
  Morel,~R.; Duraffourg,~L.; Roukes,~M.~L.; Alava,~T.; Jourdan,~G.;
  Colinet,~E.; Masselon,~C.; Brenac,~A.; Hentz,~S. Single-particle mass
  spectrometry with arrays of frequency-addressed nanomechanical resonators.
  \emph{Nat. Comm.} \textbf{2018}, \emph{9}, 3283\relax
\mciteBstWouldAddEndPuncttrue
\mciteSetBstMidEndSepPunct{\mcitedefaultmidpunct}
{\mcitedefaultendpunct}{\mcitedefaultseppunct}\relax
\EndOfBibitem
\bibitem[Barzanjeh \latin{et~al.}(2017)Barzanjeh, Wulf, Peruzzo, Kalaee,
  Dieterle, Painter, and Fink]{fink_circu}
Barzanjeh,~S.; Wulf,~M.; Peruzzo,~M.; Kalaee,~M.; Dieterle,~P.; Painter,~O.;
  Fink,~J. Mechanical on-chip microwave circulator. \emph{Nat. Comm.}
  \textbf{2017}, \emph{8}, 953\relax
\mciteBstWouldAddEndPuncttrue
\mciteSetBstMidEndSepPunct{\mcitedefaultmidpunct}
{\mcitedefaultendpunct}{\mcitedefaultseppunct}\relax
\EndOfBibitem
\bibitem[Verbridge \latin{et~al.}(2006)Verbridge, Parpia, Reichenbach, Bellan,
  and Craighead]{jeevak_APL}
Verbridge,~S.~S.; Parpia,~J.~M.; Reichenbach,~R.~B.; Bellan,~L.~M.;
  Craighead,~H.~G. High quality factor resonance at room temperature with
  nanostrings under high tensile stress. \emph{J. of Appl. Phys.}
  \textbf{2006}, \emph{99}, 124304\relax
\mciteBstWouldAddEndPuncttrue
\mciteSetBstMidEndSepPunct{\mcitedefaultmidpunct}
{\mcitedefaultendpunct}{\mcitedefaultseppunct}\relax
\EndOfBibitem
\bibitem[Verbridge \latin{et~al.}(2007)Verbridge, Shapiro, Craighead, and
  Parpia]{Jeevak_NanoLett}
Verbridge,~S.~S.; Shapiro,~D.~F.; Craighead,~H.~G.; Parpia,~J.~M. Macroscopic
  Tuning of Nanomechanics: Substrate Bending for Reversible Control of
  Frequency and Quality Factor of Nanostring Resonators. \emph{Nano Lett.}
  \textbf{2007}, \emph{7}, 1728\relax
\mciteBstWouldAddEndPuncttrue
\mciteSetBstMidEndSepPunct{\mcitedefaultmidpunct}
{\mcitedefaultendpunct}{\mcitedefaultseppunct}\relax
\EndOfBibitem
\bibitem[Fedorov \latin{et~al.}(2019)Fedorov, Engelsen, Ghadimi, Bereyhi,
  Schilling, Wilson, and Kippenberg]{PRB_kippen}
Fedorov,~S.~A.; Engelsen,~N.~J.; Ghadimi,~A.~H.; Bereyhi,~M.~J.; Schilling,~R.;
  Wilson,~D.~J.; Kippenberg,~T.~J. Generalized dissipation dilution in strained
  mechanical resonators. \emph{Phys. Rev. B} \textbf{2019}, \emph{99},
  054107\relax
\mciteBstWouldAddEndPuncttrue
\mciteSetBstMidEndSepPunct{\mcitedefaultmidpunct}
{\mcitedefaultendpunct}{\mcitedefaultseppunct}\relax
\EndOfBibitem
\bibitem[Villanueva and Schmid(2014)Villanueva, and
  Schmid]{VillanuevaSchmidPRL}
Villanueva,~L.~G.; Schmid,~S. Evidence of surface loss as ubiquitous limiting
  damping mechanism in SiN micro- and nanomechanical resonators. \emph{Phys.
  Rev. Lett.} \textbf{2014}, \emph{113}, 227201\relax
\mciteBstWouldAddEndPuncttrue
\mciteSetBstMidEndSepPunct{\mcitedefaultmidpunct}
{\mcitedefaultendpunct}{\mcitedefaultseppunct}\relax
\EndOfBibitem
\bibitem[Ftouni \latin{et~al.}(2015)Ftouni, Blanc, Tainoff, Fefferman, Defoort,
  Lulla, Richard, Collin, and Bourgeois]{FtouniPRB}
Ftouni,~H.; Blanc,~C.; Tainoff,~D.; Fefferman,~A.~D.; Defoort,~M.;
  Lulla,~K.~J.; Richard,~J.; Collin,~E.; Bourgeois,~O. Thermal conductivity of
  silicon nitride membranes is not sensitive to stress. \emph{Phys. Rev. B}
  \textbf{2015}, \emph{92}, 125439\relax
\mciteBstWouldAddEndPuncttrue
\mciteSetBstMidEndSepPunct{\mcitedefaultmidpunct}
{\mcitedefaultendpunct}{\mcitedefaultseppunct}\relax
\EndOfBibitem
\bibitem[Defoort(2014)]{martial_PHD}
Defoort,~M. \emph{Ph.D.: Non-linear dynamics in nano-electromechanical systems
  at low temperatures}; Universit\'e Grenoble Alpes: Grenoble, 2014\relax
\mciteBstWouldAddEndPuncttrue
\mciteSetBstMidEndSepPunct{\mcitedefaultmidpunct}
{\mcitedefaultendpunct}{\mcitedefaultseppunct}\relax
\EndOfBibitem
\bibitem[Hoch \latin{et~al.}(2022)Hoch, Yao, and Poot]{nano_poot}
Hoch,~D.; Yao,~X.; Poot,~M. Geometric tuning of stress in predisplaced silicon
  nitride resonators. \emph{Nano Lett.} \textbf{2022}, \emph{22}, 4013\relax
\mciteBstWouldAddEndPuncttrue
\mciteSetBstMidEndSepPunct{\mcitedefaultmidpunct}
{\mcitedefaultendpunct}{\mcitedefaultseppunct}\relax
\EndOfBibitem
\bibitem[B\"uckle \latin{et~al.}(2021)B\"uckle, Klaß, N\"agele, Braive, and
  Weig]{stress_weig}
B\"uckle,~M.; Klaß,~Y.~S.; N\"agele,~F.~B.; Braive,~R.; Weig,~E.~M. Universal
  length dependence of tensile stress in nanomechanical string resonators.
  \emph{Phys. Rev. Applied} \textbf{2021}, \emph{15}, 034063\relax
\mciteBstWouldAddEndPuncttrue
\mciteSetBstMidEndSepPunct{\mcitedefaultmidpunct}
{\mcitedefaultendpunct}{\mcitedefaultseppunct}\relax
\EndOfBibitem
\bibitem[Unterreithmeier \latin{et~al.}(2010)Unterreithmeier, Faust, and
  Kotthaus]{prl_unterrheitmeier}
Unterreithmeier,~Q.~P.; Faust,~T.; Kotthaus,~J.~P. Damping of nanomechanical
  resonators. \emph{Phys. Rev. Lett.} \textbf{2010}, \emph{105}, 027205\relax
\mciteBstWouldAddEndPuncttrue
\mciteSetBstMidEndSepPunct{\mcitedefaultmidpunct}
{\mcitedefaultendpunct}{\mcitedefaultseppunct}\relax
\EndOfBibitem
\bibitem[Yu \latin{et~al.}(2012)Yu, Purdy, and Regal]{prl_regal}
Yu,~P.-L.; Purdy,~T.~P.; Regal,~C.~A. Control of material damping in high-Q
  membrane microresonators. \emph{Phys. Rev. Lett.} \textbf{2012}, \emph{108},
  083603\relax
\mciteBstWouldAddEndPuncttrue
\mciteSetBstMidEndSepPunct{\mcitedefaultmidpunct}
{\mcitedefaultendpunct}{\mcitedefaultseppunct}\relax
\EndOfBibitem
\bibitem[Cleland(2003)]{cleland_book}
Cleland,~A. \emph{Foundations of nanomechanics}, 3rd ed.; Springer: Berlin
  Heidelberg, 2003\relax
\mciteBstWouldAddEndPuncttrue
\mciteSetBstMidEndSepPunct{\mcitedefaultmidpunct}
{\mcitedefaultendpunct}{\mcitedefaultseppunct}\relax
\EndOfBibitem
\bibitem[Olkhovets \latin{et~al.}(2000)Olkhovets, Parpia, Evoy, Carr, and
  Craighead]{jeevak_metal}
Olkhovets,~A.; Parpia,~J.~M.; Evoy,~S.; Carr,~D.; Craighead,~H.~G. Actuation
  and internal friction of torsional nanomechanical silicon resonators.
  \emph{J. Vac. Sci. Technol. B} \textbf{2000}, \emph{18}, 3549\relax
\mciteBstWouldAddEndPuncttrue
\mciteSetBstMidEndSepPunct{\mcitedefaultmidpunct}
{\mcitedefaultendpunct}{\mcitedefaultseppunct}\relax
\EndOfBibitem
\bibitem[Collin \latin{et~al.}(2010)Collin, Kofler, Lakhloufi, Pairis, Bunkov,
  and Godfrin]{eddy_JAP}
Collin,~E.; Kofler,~J.; Lakhloufi,~S.; Pairis,~S.; Bunkov,~Y.~M.; Godfrin,~H.
  Metallic coatings of microelectromechanical structures at low temperatures:
  Stress, elasticity, and nonlinear dissipation. \emph{J. of Appl. Phys.}
  \textbf{2010}, \emph{107}, 114905\relax
\mciteBstWouldAddEndPuncttrue
\mciteSetBstMidEndSepPunct{\mcitedefaultmidpunct}
{\mcitedefaultendpunct}{\mcitedefaultseppunct}\relax
\EndOfBibitem
\bibitem[Hauer \latin{et~al.}(2018)Hauer, Kim, Doolin, Souris, and
  Davis]{davisTLS}
Hauer,~B.~D.; Kim,~P.~H.; Doolin,~C.; Souris,~F.; Davis,~J.~P. Two-level system
  damping in a quasi-one dimensional optomechanical resonator. \emph{Phys. Rev.
  B} \textbf{2018}, \emph{98}, 214303\relax
\mciteBstWouldAddEndPuncttrue
\mciteSetBstMidEndSepPunct{\mcitedefaultmidpunct}
{\mcitedefaultendpunct}{\mcitedefaultseppunct}\relax
\EndOfBibitem
\bibitem[Lulla \latin{et~al.}(2013)Lulla, Defoort, Blanc, Bourgeois, and
  Collin]{kunal_PRL}
Lulla,~K.~J.; Defoort,~M.; Blanc,~C.; Bourgeois,~O.; Collin,~E. Evidence for
  the role of normal-state electrons in nanoelectromechanical damping
  mechanisms at very low temperatures. \emph{Phys. Rev. Lett.} \textbf{2013},
  \emph{110}, 177206\relax
\mciteBstWouldAddEndPuncttrue
\mciteSetBstMidEndSepPunct{\mcitedefaultmidpunct}
{\mcitedefaultendpunct}{\mcitedefaultseppunct}\relax
\EndOfBibitem
\bibitem[Maillet \latin{et~al.}(2023)Maillet, Cattiaux, Zhou, Gazizulin,
  Bourgeois, Fefferman, and Collin]{OliveArXiv}
Maillet,~O.; Cattiaux,~D.; Zhou,~X.; Gazizulin,~R.~R.; Bourgeois,~O.;
  Fefferman,~A.~D.; Collin,~E. Nanomechanical damping via electron-assisted
  relaxation of two-level systems. \emph{Phys. Rev. B} \textbf{2023},
  \emph{107}, 064104\relax
\mciteBstWouldAddEndPuncttrue
\mciteSetBstMidEndSepPunct{\mcitedefaultmidpunct}
{\mcitedefaultendpunct}{\mcitedefaultseppunct}\relax
\EndOfBibitem
\bibitem[Photiadis and Judge(2004)Photiadis, and Judge]{judge1}
Photiadis,~D.~M.; Judge,~J.~A. Attachment losses of high Q oscillators.
  \emph{Appl. Phys. Lett.} \textbf{2004}, \emph{85}, 482\relax
\mciteBstWouldAddEndPuncttrue
\mciteSetBstMidEndSepPunct{\mcitedefaultmidpunct}
{\mcitedefaultendpunct}{\mcitedefaultseppunct}\relax
\EndOfBibitem
\bibitem[Judge \latin{et~al.}(2007)Judge, Photiadis, Vignola, Houston, and
  Jarzynski]{judge2}
Judge,~J.~A.; Photiadis,~D.~M.; Vignola,~J.~F.; Houston,~B.~H.; Jarzynski,~J.
  Attachment loss of micromechanical and nanomechanical resonators in the
  limits of thick and thin support structures. \emph{J. of Appl. Phys.}
  \textbf{2007}, \emph{101}, 013521\relax
\mciteBstWouldAddEndPuncttrue
\mciteSetBstMidEndSepPunct{\mcitedefaultmidpunct}
{\mcitedefaultendpunct}{\mcitedefaultseppunct}\relax
\EndOfBibitem
\bibitem[Cross and Lifshitz(2001)Cross, and Lifshitz]{crossPRB}
Cross,~M.~C.; Lifshitz,~R. Elastic wave transmission at an abrupt junction in a
  thin plate with application to heat transport and vibrations in mesoscopic
  systems. \emph{Phys. Rev. B} \textbf{2001}, \emph{64}, 085324\relax
\mciteBstWouldAddEndPuncttrue
\mciteSetBstMidEndSepPunct{\mcitedefaultmidpunct}
{\mcitedefaultendpunct}{\mcitedefaultseppunct}\relax
\EndOfBibitem
\bibitem[Wilson-Rae(2008)]{ignacio}
Wilson-Rae,~I. Intrinsic dissipation in nanomechanical resonators due to phonon
  tunneling. \emph{Phys. Rev. B} \textbf{2008}, \emph{77}, 245418\relax
\mciteBstWouldAddEndPuncttrue
\mciteSetBstMidEndSepPunct{\mcitedefaultmidpunct}
{\mcitedefaultendpunct}{\mcitedefaultseppunct}\relax
\EndOfBibitem
\bibitem[Schmid \latin{et~al.}(2011)Schmid, Jensen, Nielsen, and
  Boisen]{schmidPRB2011}
Schmid,~S.; Jensen,~K.~D.; Nielsen,~K.~H.; Boisen,~A. Damping mechanisms in
  high-Q micro and nanomechanical string resonators. \emph{Phys. Rev. B}
  \textbf{2011}, \emph{84}, 165307\relax
\mciteBstWouldAddEndPuncttrue
\mciteSetBstMidEndSepPunct{\mcitedefaultmidpunct}
{\mcitedefaultendpunct}{\mcitedefaultseppunct}\relax
\EndOfBibitem
\bibitem[Ghadimi \latin{et~al.}(2017)Ghadimi, Wilson, and
  Kippenberg]{kippNanoLett}
Ghadimi,~A.~H.; Wilson,~D.~J.; Kippenberg,~T.~J. Radiation and internal loss
  engineering of high-stress silicon nitride nanobeams. \emph{Nano Lett.}
  \textbf{2017}, \emph{17}, 3501\relax
\mciteBstWouldAddEndPuncttrue
\mciteSetBstMidEndSepPunct{\mcitedefaultmidpunct}
{\mcitedefaultendpunct}{\mcitedefaultseppunct}\relax
\EndOfBibitem
\bibitem[Adiga \latin{et~al.}(2012)Adiga, Ilic, Barton, Wilson-Rae, Craighead,
  and Parpia]{JeevakMaxQ}
Adiga,~V.~P.; Ilic,~B.; Barton,~R.~A.; Wilson-Rae,~I.; Craighead,~H.~G.;
  Parpia,~J.~M. Approaching intrinsic performance in ultra-thin silicon nitride
  drum resonators. \emph{J. of Appl. Phys.} \textbf{2012}, \emph{112},
  064323\relax
\mciteBstWouldAddEndPuncttrue
\mciteSetBstMidEndSepPunct{\mcitedefaultmidpunct}
{\mcitedefaultendpunct}{\mcitedefaultseppunct}\relax
\EndOfBibitem
\bibitem[Tsaturyan \latin{et~al.}(2017)Tsaturyan, Barg, Polzik, and
  Schliesser]{schliesser_Q}
Tsaturyan,~Y.; Barg,~A.; Polzik,~E.~S.; Schliesser,~A. Ultracoherent
  nanomechanical resonators via soft clamping and dissipation dilution.
  \emph{Nat. Nanotech.} \textbf{2017}, \emph{12}, 776\relax
\mciteBstWouldAddEndPuncttrue
\mciteSetBstMidEndSepPunct{\mcitedefaultmidpunct}
{\mcitedefaultendpunct}{\mcitedefaultseppunct}\relax
\EndOfBibitem
\bibitem[Yu \latin{et~al.}(2014)Yu, Cicak, Kampel, Tsaturyan, Purdy, Simmonds,
  and Regal]{RegalAPL}
Yu,~P.-L.; Cicak,~K.; Kampel,~N.~S.; Tsaturyan,~Y.; Purdy,~T.~P.;
  Simmonds,~R.~W.; Regal,~C.~A. A phononic bandgap shield for high-Q membrane
  microresonators. \emph{Appl. Phys. Lett.} \textbf{2014}, \emph{104},
  023510\relax
\mciteBstWouldAddEndPuncttrue
\mciteSetBstMidEndSepPunct{\mcitedefaultmidpunct}
{\mcitedefaultendpunct}{\mcitedefaultseppunct}\relax
\EndOfBibitem
\bibitem[Suhel \latin{et~al.}(2012)Suhel, Hauer, Biswas, Beach, and
  Davis]{SuhelPRB}
Suhel,~A.; Hauer,~B.~D.; Biswas,~T.~S.; Beach,~K. S.~D.; Davis,~J.~P.
  Dissipation mechanisms in thermomechanically driven silicon nitride
  nanostrings. \emph{Appl. Phys. Lett.} \textbf{2012}, \emph{100}, 173111\relax
\mciteBstWouldAddEndPuncttrue
\mciteSetBstMidEndSepPunct{\mcitedefaultmidpunct}
{\mcitedefaultendpunct}{\mcitedefaultseppunct}\relax
\EndOfBibitem
\bibitem[Ghadimi \latin{et~al.}(2018)Ghadimi, Fedorov, Engelsen, Bereyhi,
  Schilling, Wilson, and Kippenberg]{kipp_science}
Ghadimi,~A.~H.; Fedorov,~S.~A.; Engelsen,~N.~J.; Bereyhi,~M.~J.; Schilling,~R.;
  Wilson,~D.~J.; Kippenberg,~T.~J. Elastic strain engineering for ultralow
  mechanical dissipation. \emph{Science} \textbf{2018}, \emph{360}, 764\relax
\mciteBstWouldAddEndPuncttrue
\mciteSetBstMidEndSepPunct{\mcitedefaultmidpunct}
{\mcitedefaultendpunct}{\mcitedefaultseppunct}\relax
\EndOfBibitem
\bibitem[Sadeghi \latin{et~al.}(2019)Sadeghi, Tanzer, Christensen, and
  Schmid]{schmidJAProundclamp}
Sadeghi,~P.; Tanzer,~M.; Christensen,~S.~L.; Schmid,~S. Influence of
  clamp-widening on the quality factor of nanomechanical silicon nitride
  resonators. \emph{J. of Appl. Phys.} \textbf{2019}, \emph{126}, 165108\relax
\mciteBstWouldAddEndPuncttrue
\mciteSetBstMidEndSepPunct{\mcitedefaultmidpunct}
{\mcitedefaultendpunct}{\mcitedefaultseppunct}\relax
\EndOfBibitem
\bibitem[Babaei~Gavan \latin{et~al.}(2009)Babaei~Gavan, van~der Drift, Venstra,
  Zuiddam, and van~der Zant]{VdZ_freq}
Babaei~Gavan,~K.; van~der Drift,~E. W. J.~M.; Venstra,~W.~J.; Zuiddam,~M.~R.;
  van~der Zant,~H. S.~J. Effect of undercut on the resonant behaviour of
  silicon nitride cantilevers. \emph{J. of Micromech. Microeng.} \textbf{2009},
  \emph{19}, 035003\relax
\mciteBstWouldAddEndPuncttrue
\mciteSetBstMidEndSepPunct{\mcitedefaultmidpunct}
{\mcitedefaultendpunct}{\mcitedefaultseppunct}\relax
\EndOfBibitem
\bibitem[Golokolenov \latin{et~al.}(2023)Golokolenov, Alperin, Fernandez,
  Fefferman, and Collin]{Ilya_JLTP}
Golokolenov,~I.; Alperin,~B.; Fernandez,~B.; Fefferman,~A.; Collin,~E. Fully
  Suspended Nano-beams for Quantum Fluids. \emph{J. of Low Temp. Phys.}
  \textbf{2023}, \emph{210}, 550–561\relax
\mciteBstWouldAddEndPuncttrue
\mciteSetBstMidEndSepPunct{\mcitedefaultmidpunct}
{\mcitedefaultendpunct}{\mcitedefaultseppunct}\relax
\EndOfBibitem
\bibitem[Cleland and Roukes(1999)Cleland, and Roukes]{roukesclelandsensors}
Cleland,~A.; Roukes,~M. External control of dissipation in a nanometer-scale
  radiofrequency mechanical resonator. \emph{Sensors and Actuators}
  \textbf{1999}, \emph{72}, 256\relax
\mciteBstWouldAddEndPuncttrue
\mciteSetBstMidEndSepPunct{\mcitedefaultmidpunct}
{\mcitedefaultendpunct}{\mcitedefaultseppunct}\relax
\EndOfBibitem
\bibitem[Timoshenko \latin{et~al.}(1974)Timoshenko, Young, and
  Weaver]{timoshenko}
Timoshenko,~S.; Young,~D.; Weaver,~W. \emph{Vibrations problems in
  engineering}, 4th ed.; John Wiley and Sons, 1974\relax
\mciteBstWouldAddEndPuncttrue
\mciteSetBstMidEndSepPunct{\mcitedefaultmidpunct}
{\mcitedefaultendpunct}{\mcitedefaultseppunct}\relax
\EndOfBibitem
\bibitem[Bokaian(1990)]{bokaian}
Bokaian,~A. Natural frequencies of beams under tensile axial loads. \emph{J. of
  Sound and Vibr.} \textbf{1990}, \emph{142}, 481\relax
\mciteBstWouldAddEndPuncttrue
\mciteSetBstMidEndSepPunct{\mcitedefaultmidpunct}
{\mcitedefaultendpunct}{\mcitedefaultseppunct}\relax
\EndOfBibitem
\bibitem[Sadewasser \latin{et~al.}(2006)Sadewasser, Villanueva, and
  Plaza]{Sadewasser}
Sadewasser,~S.; Villanueva,~G.; Plaza,~J.~A. Modified atomic force microscopy
  cantilever design to facilitate access of higher modes of oscillation.
  \emph{Review of Sci. Instruments} \textbf{2006}, \emph{77}, 073703\relax
\mciteBstWouldAddEndPuncttrue
\mciteSetBstMidEndSepPunct{\mcitedefaultmidpunct}
{\mcitedefaultendpunct}{\mcitedefaultseppunct}\relax
\EndOfBibitem
\bibitem[Zener(1948)]{zenerbook}
Zener,~C. \emph{Elasticity and Anelasticity of Metals}, 4th ed.; The University
  of Chicago Press, 1948\relax
\mciteBstWouldAddEndPuncttrue
\mciteSetBstMidEndSepPunct{\mcitedefaultmidpunct}
{\mcitedefaultendpunct}{\mcitedefaultseppunct}\relax
\EndOfBibitem
\bibitem[Landau and Lifshitz(1987)Landau, and Lifshitz]{landaufluids}
Landau,~L.; Lifshitz,~E. \emph{Fluid mechanics}, 2nd ed.; Pergamon Press:
  Headington Hill Hall, 1987\relax
\mciteBstWouldAddEndPuncttrue
\mciteSetBstMidEndSepPunct{\mcitedefaultmidpunct}
{\mcitedefaultendpunct}{\mcitedefaultseppunct}\relax
\EndOfBibitem
\bibitem[Collin(2022)]{cloud}
Collin,~E. Data for Nano beam Clamping Revisited. \emph{Golokolenov2022 N\'eel}
  \textbf{2022}, \relax
\mciteBstWouldAddEndPunctfalse
\mciteSetBstMidEndSepPunct{\mcitedefaultmidpunct}
{}{\mcitedefaultseppunct}\relax
\EndOfBibitem
\end{mcitethebibliography}

\end{document}